\documentclass[preprint,showpacs,preprintnumbers,amsmath,amssymb, superscriptaddress,longbibliography,nofootinbib]{revtex4-1}
\usepackage{graphicx} 
\usepackage{xcolor}
\usepackage{amsmath}
 \usepackage{hyperref}
\usepackage[top = 2cm, bottom = 2cm, right = 2cm, left =2cm]{geometry}

\begin{document}

\title{A way of decoupling the gravitational bulk field equations of {regular} braneworld black holes to suppress the bulk singularities.}

\author{Milko Estrada \footnote{E-mail: milko.estrada@gmail.com}}\affiliation{Facultad de Ingeniería y Empresa, Universidad Católica Silva Henríquez, Chile}

\author{T. M. Crispim \footnote{E-mail: iago.crispim@fisica.ufc.br}}\affiliation{Departamento de F\'isica, Universidade Federal do Cear\'a, Caixa Postal 6030, Campus do Pici, 60455-760 Fortaleza, Cear\'a, Brazil}

\author{G. Alencar \footnote{E-mail: eova@fisica.ufc.br}}\affiliation{Departamento de F\'isica, Universidade Federal do Cear\'a, Caixa Postal 6030, Campus do Pici, 60455-760 Fortaleza, Cear\'a, Brazil}

\date{\today}

\begin{abstract}
We provide a methodology for decoupling the bulk gravitational field equations of braneworld black holes to suppress the bulk singularities. {Thus, we provide a regular braneworld black hole setup}. To achieve this, we apply a  Minimal Geometric Deformation (MGD) with respect to a coupling constant \( \alpha \) to the $4D$ Minkowski spacetime embedded in an extra dimension. This results in a gravitational decoupling into a system \( \mathcal{A} \) with equations of motion of order \( \alpha^0 \) and a system \( \mathcal{B} \), related to the so-called Quasi-Einstein equations of order \( \alpha \). This methodology allows for the construction of a regular geometry everywhere. We outline the necessary constraints for eliminating singularities and provide a recipe for solving the equations of motion. Both the warp factor, the scalar field, and the potential obtained are smooth and free from Dirac delta singularities. A control parameter is introduced such that, in the limit \( b \to 0 \), the Randall-Sundrum (RS) setup is recovered, resulting in a transition from a thick brane to a thin brane. The asymptotic behavior of the curvature invariant \( \displaystyle \lim_{y \to \pm \infty} R_{5D}(r,y) \) is positive near the de Sitter core (for small \( r \)), asymptotically negative for finite \( r > r_* \), and asymptotically flat at the $4D$ boundary as \( r \to \infty \). Although this work aims to suppress bulk singularities, it is expected that our methodology may be useful for future investigations related to the embedding of gravitational objects within other braneworld contexts.
\end{abstract}

\maketitle
\section{introduction}

The recent detection of gravitational waves by LIGO and VIRGO, coupled with the first image of a supermassive black hole captured by the Event Horizon Telescope (EHT) \cite{LIGOScientific:2016aoc, LIGOScientific:2017vwq, EventHorizonTelescope:2022wkp}, has provided experimental data on gravity in the universe. These experiments have allowed us to test gravitational theories beyond General Relativity. In this context, braneworld theories have been tested using data from these experiments in recent years \cite{Banerjee:2021aln, Banerjee:2022jog, Visinelli:2017bny,Mishra:2021waw}. However, it is worth mentioning that the analyses conducted in these recent references seem not to consider the LHC results. {Regarding this, some efforts to compare brane-world models with LHC results have been made in the literature. In reference \cite{Casadio:2012pu}, the minimal geometric deformation of stars in the brane-world is studied, and brane-world black hole metrics are found. It is stated that a minimum mass for semiclassical microscopic black holes can be derived, which has a relevant impact on the description of black hole events at the LHC. Also, see reference \cite{daRocha:2006ei}, where the possible effects of the brane world are investigated for mini-black holes and their measurable consequences in Large Hadron Collider observations. It is stated that if the fundamental Planck scale is at the TeV scale, the LHC will produce more than $10^7$ black holes per year. Therefore, contrasting brane-world models with experimental results seems to be an open issue. See more recent studies for example in references \cite{Das:2020gie,Belis:2023atb}} The Randall-Sundrum (RS) model \cite{Randall:1999vf,Randall:1999ee} offers a geometric explanation for why gravity is weaker than the other fundamental forces of nature.

The RS model involves the embedding of four-dimensional flat branes within a five-dimensional anti-de Sitter (AdS) spacetime. On the other hand, since the gravitational collapse of a black hole is described by the Schwarzschild metric, it is natural to study this $4D$ geometry localized on a brane embedded in an extra dimension, that is, in a brane-world setup. This problem was initially investigated in reference \cite{Chamblin:1999by}, specifically addressing the so-called black string. However, when embedding a Schwarzschild black hole in an extra dimension within the Randall-Sundrum framework, physical singularities arise both at the origin of the radial coordinate and at the infinity of the extra dimension. This can be tested by evaluating the Kretschmann scalar.
\begin{equation} \label{RicciScalarSch}
     K_{5D} \sim 40k^4 + 16 M^2 \frac{e^{4ky}}{ r^6}
\end{equation}

It is also important to mention that the confinement of \(3\)-branes implies the presence of a pathology in the equations of motion at the brane locations, characterized by Dirac delta-type singularities. One way to address this pathology associated with the Dirac delta in the action principle of a \(3\)-brane was proposed in reference \cite{Kehagias:2000au}. This approach consists of replacing the Dirac delta with a scalar field \(\phi\) that resides in this space and is coupled through the standard action. Notably, in this model, \(\phi\)  physically represents the material from which the brane is constructed. Thus, resulting in a thick brane configuration. The warp factor \(A(y)\) is defined to ensure that its value, as well as its first and second derivatives evaluated at the center of the extra coordinate domain (\(y = 0\)), varies smoothly and takes on finite values. This contrasts with the Dirac delta-type discontinuities present in the Randall-Sundrum model. Furthermore, in these models, a control parameter \(b\) is introduced, which establishes a specific limit where the geometry and matter configuration recover the Randall-Sundrum framework. That is to say, a transition occurs from a thick brane to the thin brane of the RS model. See examples of recent applications in \cite{Hendi:2020qkk,Peyravi:2022ubf,Rosa:2022fhl}.

It is worth noting that the approaches to studying the representation of black holes (BHs) in a braneworld setup are primarily based on two methodologies: one originating from the brane and the other from the bulk. In the former approach, four-dimensional black hole solutions are constructed by studying the equations of motion induced on the brane \cite{Shiromizu:1999wj}. Some attempts to suppress the central radial singularity using 4D RBHs can be found in reference \cite{Neves:2021dqx}. However, this methodology does not fully capture the influence of the bulk geometry on the physical and geometric behavior of the four-dimensional spacetime, and, among other limitations, it does not allow for a complete analysis of singularities in the extra dimension. On the other hand, the second approach constructs spacetimes directly from the bulk, with the brane equations derived from there. This work will focus on this latter perspective. See for example \cite{Chamblin:1999by,Kanti:2002fx,Hirayama:2001bi,Kanti:2003uv,Kanti:2003uv} and more recent applications in \cite{Nakas:2020sey,Nakas:2021srr,Nakas:2023yhj,Crispim:2024nou,Crispim:2024yjz}. 

On the other hand, in recent years, the {\it Gravitational Decoupling} (GD) algorithm \cite{Ovalle:2017fgl} has garnered significant attention for yielding new solutions of physical interest. To name just a few cases, this method has been used to generate rotating black holes \cite{Contreras:2021yxe}, hairy black holes \cite{Ovalle:2020kpd}, spherically symmetric black holes \cite{Ovalle:2018umz}, regular black holes \cite{Ovalle:2023ref}, stellar distributions \cite{Estrada:2018zbh,Estrada:2018vrl}, Charged Anisotropic Spherical Solutions \cite{Sharif:2018toc}, solutions in Lovelock gravity \cite{Estrada:2019aeh}, among others. This method was also studied for braneworld models in reference \cite{Leon:2019abq}, but using the aforementioned methodology, based on the induced equations of motion on the brane. {Broadly speaking, this method is explained in reference \cite{Ovalle:2017wqi} as follows: ``given two gravitational sources $ \mathcal{A} $ and $ \mathcal{B} $, standard Einstein's equations are first solved for $ \mathcal{A} $, and then a simpler set of quasi-Einstein equations are solved for $ \mathcal{B} $. Finally, the two solutions can be combined in order to derive the complete solution for the total system $ \mathcal{A} \cup \mathcal{B} $." It is worth mentioning that this method is not perturbative \cite{Sharif:2020llo,daRocha:2021sqd}, but rather based on an exact gravitational decoupling.}

Since the method mentioned in the previous paragraph (GD) has been used to test various gravitational scenarios, and given the growing interest in recent years to test braneworld models with experimental data \cite{Banerjee:2021aln, Banerjee:2022jog, Visinelli:2017bny,Mishra:2021waw}, it would be quite beneficial to provide a methodology analogous to GD from the bulk perspective. This would enable the exploration of different gravitational scenarios within a braneworld setup.

In this work, we present a methodology to decouple the equations of motion for braneworld models, using the approach based on the bulk equations of motion. The methodology introduced here is analogous but distinct from the one mentioned previously, corresponding to reference \cite{Ovalle:2017fgl}. In this work, the proposed methodology aims to eliminate the singularity at infinity in the extra coordinate generated by embedding a black hole within a braneworld framework. To achieve this, we will initially consider a gravitational system \( \mathcal{A} \) associated with the $4D$ Minkowski seed spacetime embedded in an extra dimension, along with matter sources given by the scalar field and its potential. We will then linearly deform the 4D Minkowski geometry with respect to the coupling constant \( \alpha \). This geometric deformation will be coupled to a bulk energy-momentum tensor of order \( \alpha \), giving rise to a new gravitational system \( \mathcal{B} \), whose equations of motion, as we will see below, are also of order \( \alpha \). These equations are analogous to the so-called quasi-Einstein equations. Consequently, through the equations of order \( \alpha \), we will suppress the singularity at the radial origin, while the equations of order \( \alpha^0 \) will handle the singularity at infinity in the extra dimension. We will outline the constraints that must be satisfied to eliminate the singularity present both at the origin of the radial coordinate and at infinity in the extra coordinate. Additionally, we will provide a recipe for solving our equations of motion. Furthermore, we will identify the limits at which a thick brane in the bulk transitions to a thin brane and analyze the physical characteristics of our model. 

Although this work presents a methodology aimed at eliminating the singularity at infinity, it is expected that the provided methodology will serve as a basis for studying other future problems related to the embedding of gravitational objects in a braneworld setup. For more complex scenarios, this would likely require redefining the coupling constants and the matter sources coupled in the bulk.

\subsection{The geometry and Invariants of curvature} \label{SeccionInvariantes}

We will consider the following line element:
\begin{equation} \label{NuestroElementoDeLinea}
ds^2=e^{2A(y)}\hat{g}_{\mu\nu}dx^{\mu}dx^{\nu}+dy^2
\end{equation}
where
\begin{equation} \label{elementoDeLinea4D}
\hat{g}_{\mu\nu}dx^{\mu}dx^{\nu} = - U(r) dt^2 + \frac{dr^2}{U(r)} + r^2 d\Omega^2   
\end{equation}
where $x^\mu=(t,r,\theta,\phi)$.

For the line element \eqref{NuestroElementoDeLinea}, the corresponding expression for the Ricci scalar is:
\begin{align} \label{Ricci5D}
    R_{5D}=-20 ({A}')^2-8 {A}''+ e^{-2A(y)}\bar{R}_{4D},
\end{align}
where the $'$ indicates derivation with respect to the extra coordinate ($z$), and where  $\bar{R}_{4D}$ corresponds to the Ricci scalar of the four dimensional geometry whose line element is given by equation \eqref{elementoDeLinea4D}:
\begin{equation} \label{Ricci4D}
    \bar{R}_{4D}= -\frac{\partial^2 U}{\partial r^2}- \frac{4}{r}\frac{\partial U}{\partial r}+ \frac{2}{r^2} \left ( 1-U \right) .
\end{equation}

On the other hand, the Kretschmann scalar is given by:
\begin{align} \label{Kre5D}
    K_{5D}= 40 ({A'})^4+16 ({A''})^2+32 {A''} ({A'})^2- 4 e^{-2A(y)} ({A'})^2 \bar{R}_{4D} + e^{-4A(y)} \bar{K}_{4D},
\end{align}
where the four dimensional Kretschmann scalar $\bar{K}_{4D}$ corresponds to the four dimenional geometry, equation \eqref{elementoDeLinea4D}
\begin{equation} \label{4D}
   \bar{K}_{4D}=\left (\frac{\partial^2 U}{\partial r^2} \right )^2+ \frac{4}{r^2} \left (\frac{\partial U}{\partial r} \right )^2 + \frac{4}{r^4} (1-U)^2 .
\end{equation}

Below, we will present a list of criteria that must be satisfied for our geometry to be considered regular. These criteria point out that both the curvature invariants and the equations of motion (which we will test below) must be regular everywhere.

\section{Our propose}

\subsection{Action and equations of motion}
Our action is determined by
\begin{equation}\label{AccionInicialSuave}
S = \int d^4x\int dy \sqrt{-g}\left(2M^3R - \frac{1}{2}(\partial\phi)^2 - V(\phi) - \alpha \cdot L_B   \right)
\end{equation}
where \(\alpha\) is a dimensionless constant and where $L_B$ is the bulk lagrangian \cite{Kanti:2001cj,Binetruy:1999ut}. Further below, we will explore the role that the constant \(\alpha\) plays in the equations of motion. 

Applying the variational principle to this action \eqref{AccionInicialSuave}, we obtain the following set of equations:
\begin{equation}\label{eqMotionGral}
    R_{MN}-\dfrac{1}{2}g_{MN}R = \dfrac{1}{4M^{3}}\biggl( \partial_{M}\phi\partial_{N}\phi - g_{MN}\biggl[\dfrac{1}{2}(\partial\phi)^{2} + V(\phi) \biggl] - \alpha \cdot T_{MN}^{{(bulk)}} \biggl),
\end{equation}

\begin{equation}\label{eqMotionV}
    \dfrac{1}{\sqrt{-g}}\partial_{M}(\sqrt{-g} g^{MN}\partial_{N}\phi)=\dfrac{\partial V(\phi)}{\partial \phi},
\end{equation}

\begin{equation} \label{ConservacionBulk}
    \nabla_{M} T^{MN}_{(bulk)} = 0
\end{equation}
where $T^{MN}_{(bulk)}$ corresponds to the stress--energy tensor for bulk matter.  As we will discuss below, for our geometric distribution, the corresponding energy-momentum tensor in the bulk must be diagonal, which we will write as: 

\begin{equation} \label{EMBulk}
    (T^M_N)_{(bulk)}(r,y) = \mbox{diag} (-\rho,p_r,p_t,p_t, T^y_y)
\end{equation}

The conservation equation \eqref{ConservacionBulk} gives rise to:

\begin{align}
    \frac{\partial}{\partial r} p_r + \frac{2}{r} \left ( p_r- p_t \right ) &=0 \label{ConservacionR} \\
    \frac{\partial T^y_y}{\partial y}  + \frac{\partial A}{\partial y} \left ( 4 T^y_y - 2p_r - 2 p_t \right )&=0 \label{ConservacionY}
\end{align}

\subsection{Gravitational Decoupling--like background for the bulk}

To explain our methodology, in this section we will highlight some aspects of our Gravitational Decoupling-like background for the bulk.

\begin{enumerate}
    \item Initially, we consider a gravitational system $\mathcal{A}$, whose geometry and matter source are of order \( \alpha^0 \). That is, for this system we consider the coupling constant \( \alpha \) to be turned off, thus, $\alpha=0$. 
    \begin{itemize}
        \item The matter sources for $\mathcal{A}$ are associated with the second and third terms in action \eqref{AccionInicialSuave}, which are related to the scalar field and the potential, respectively.
        \item  The geometry of $\mathcal{A}$ corresponds to a seed spacetime where  $(\hat{g}_{\mu \nu})_{seed} = \eta_{\mu \nu}$ , i.e., \( U(r)_{seed} = 1 \), representing a Minkowski space embedded in an extra dimension, as given by the line element in equation\eqref{NuestroElementoDeLinea}.
    \end{itemize}

As we will see below, the equations of motion at order \( \alpha^0 \), i.e., with the coupling constant \( \alpha \) turned off, correspond to equations  \eqref{ComponenteYAzul}, \eqref{ComponenteTAzul}  and \eqref{eqMotionV}.
    
    \item We now proceed to turn on the coupling constant \( \alpha \), giving rise to the gravitational system $\mathcal{B}$, as follows:
    \begin{itemize}
        \item The seed geometry is linearly deformed with respect to a dimensionless coupling constant \( \alpha \) as follows
        \begin{equation} \label{MGD1}
   U(r)_{seed} =1 \to U(r)= 1-\alpha \cdot \frac{m(r)}{r} \Rightarrow ({g}_{\mu \nu})_{seed} = \eta_{\mu \nu} \to \hat{g}_{\mu \nu}
\end{equation}
Thus, the function \( m(r)/r \) represents a {\it minimal geometric deformation} (MGD) \cite{Ovalle:2017fgl} of Minkowski spacetime, thus, $\displaystyle \lim_{\alpha \to 0} \hat{g}_{\mu\nu} = \eta_{\mu\nu}$.
That is, turning on the constant \( \alpha \), due to the minimal geometric deformation, \( U(r)_{seed} \) becomes \( U(r) \), and for our case, \( ({g}_{\mu \nu})_{seed} = \eta_{\mu \nu} \) transitions to \( \hat{g}_{\mu \nu} \). 
\item This geometric deformation is coupled to a bulk matter source, which is also of order \( \alpha \), as can be seen in the last term of equation \eqref{eqMotionGral}. This gives rise to a gravitational system $\mathcal{B}$, whose equations of motion, as we will see below, are of order \( \alpha \). These equations are analogous to the so-called quasi-Einstein equations \cite{Ovalle:2017fgl}.
    \end{itemize}
\item As we will see below, as a consequence of the previous steps, the system of equations of motion will be decoupled into two simpler systems: the system $\mathcal{A}$, whose equations of motion are of order \( \alpha^0 \), and the system $\mathcal{B}$, whose equations of motion, the so--called Quasi Einstein equations, are of order \( \alpha \).
\end{enumerate}

\subsection{Equations of motion}

As mentioned in the introduction, this work provides a way to decouple the equations of motion in order to eliminate the singularity present at infinity in the radial coordinate. In this way, the constant \(\alpha\) allows the decoupling of the equations into a part that depends on \((r, y)\) and a part that depends solely on the coordinate \((y)\). This form allows us to solve each component of the equations of motion separately. Below, we will describe why this way of decoupling will lead to a convenient way to find a solution corresponding to the coupled system.

We impose, for simplicity, that $\phi = \phi(y)$ and that $V(\phi) =V(\phi(y))$.

As we can see, the equations of motion split in the following manner:

\subsubsection{\bf (y) dependent equations of $\alpha^0$ order} 

Following the indications from point 1 of the previous section, we initially set the coupling constant \( \alpha \) turned off, using as embedded seed spacetime  \( (\hat{g}_{\mu \nu})_{seed} = \eta_{\mu \nu} \), i.e., \( U(r)_{seed} = 1 \), which represents a Minkowski space embedded in an extra dimension. The matter sources are provided by the scalar field and the potential. Thus, the $y$ dependent equations of motion of order \( \alpha^0 \) are:

\begin{align} \label{ComponenteYAzul}
 -\frac{1}{2} \left(\frac{\partial \phi}{\partial y}\right)^2 + V(\phi) = -24 M^3 \left(\frac{\partial A}{\partial y}\right)^2 ,
\end{align}

\begin{align} \label{ComponenteTAzul}
    \frac{1}{2} \left(\frac{\partial \phi}{\partial y}\right)^2 + V(\phi) = -12 M^3 \left(\frac{\partial^2 A}{\partial y^2}\right) - 24 M^3 \left(\frac{\partial A}{\partial y}\right)^2
\end{align}

The conservation equation associated with the matter sources of order \( \alpha^0 \), i.e., the scalar field and the potential, is given by equation \eqref{eqMotionV}. Since two of these three equations are independent \cite{Hendi:2020qkk}, due to the Bianchi identities, it is not necessary to develop the latter to obtain the solution.

\subsubsection{\bf Quasi Einstein (r,y) dependent equations of $\alpha$ order}

By turning on the constant \( \alpha \) and following the indications from point 2 of the previous section, we obtain the quasi-Einstein equations dependent on \( (r, y) \) at order \( \alpha \).

\begin{align} \label{ComponenteTNegro}
     p_r=\frac{8 e^{-2 A(y)}}{r^2} M^3 \frac{\partial m}{\partial r}
\end{align}

\begin{align} \label{ComponenteThetaNegro}
 p_t= \frac{4 e^{-2 A(y)} M^3 \frac{\partial^2 m}{\partial r^2}}{r}
\end{align}
and
\begin{align} \label{ComponenteYNegro}
  T^y_y= \frac{8 e^{-2 A(y)}}{r^2} M^3 \frac{\partial m}{\partial r}+ \frac{4 e^{-2 A(y)}}{r} M^3 \frac{\partial^2 m}{\partial r^2}
\end{align}

Comparing equations \eqref{ComponenteTNegro}, \eqref{ComponenteThetaNegro}, and \eqref{ComponenteYNegro}, it is evident that
\begin{equation} \label{SumaPresiones}
   T^y_y= p_r+ p_t
\end{equation}

However, this must be consistent with equation \eqref{ConservacionY}.Thus, by introducing condition \eqref{SumaPresiones} into equation \eqref{ConservacionY}:
\begin{equation} \label{ComponenteEMY}
    T^y_y = e^{-2A(y)} \bar{T}^y_y(r)
\end{equation}

In this way, it is straightforward to verify that the energy-momentum tensor in the bulk has the following structure:
\begin{equation} \label{EMBulkSplit}
    (T^M_N)_{(bulk)} =  e^{-2A(y)} \mbox{diag} \left (-\bar{\rho}(r),\bar{p}_r(r),\bar{p}_t(r),\bar{p}_t(r), \bar{T}^y_y(r) \right) = e^{-2A(y)} \bar{T}^M_N (r)
\end{equation}

Thus, for the imposed five-dimensional geometry, the trace condition from reference \cite{Alencar:2024lrl} must be satisfied, namely:
\begin{equation} \label{ConservacionYCuasi}
2 T^y_y=T^\mu_\mu \Rightarrow 2 \bar{T}^y_y=\bar{T}^\mu_\mu 
\end{equation}

On the other hand, the radial component of the bulk conservation equation \eqref{ConservacionR} reduces to:
\begin{equation} \label{ConservacionRCuasi}
    \frac{\partial}{\partial r} \bar{p}_r + \frac{2}{r} \left ( \bar{p}_r- \bar{p}_t \right ) =0 
\end{equation}

It is worth stressing that the fact that the conservation equation of the system at order \( \alpha^0 \) (equation \ref{eqMotionV}) is independent of the conservation equations of the Quasi-Einstein equations at order \( \alpha \) (equations \ref{ConservacionYCuasi} and \ref{ConservacionRCuasi}) implies that there is no exchange of energy-momentum between the fluid of the seed geometry (given by the scalar field and the potential) and the matter fields of the bulk. Therefore, there is only a purely gravitational interaction between both systems, $\mathcal{A}$ and $\mathcal{B}$. This is a common feature of the gravitational decoupling method.

Below, we will outline a list of criteria to ensure that the geometry is regular.

\subsection{A list of criteria for ensuring the regularity of the geometry}

Related to the curvature invariants:

\begin{enumerate}
    \item \label{CondicionRho}  In order to eliminate the singularity located at the radial origin, the embedded geometry must be regular everywhere. In this work, the embedded geometry corresponds to a Regular Black Hole (RBH). Thus, both \(\bar{R}_{4D}\) and \(\bar{K}_{4D}\) must be finite at every location.

Although various types of RBH exist in the literature, for the sake of simplicity, this work will focus on the embedding of a 4D RBH with a de Sitter core. Regarding this point, it is important to note that finding a $4D$ geometry $\hat{g}_{\mu\nu}$ that satisfies this condition is not novel in the literature, as there is an extensive range of geometries representing $4D$ regular black holes (RBHs). Some examples in references \cite{Hayward:2005gi,Dymnikova:1992ux,Spallucci:2017aod}.  The main characteristics of RBHs with a de Sitter core that will be useful for this work are presented in Appendix \ref{RBH4D}.

However, as mentioned, part of the novelty of this work lies in our goal to establish a consistent framework for solving the bulk equations of motion and the bulk energy-momentum tensor through the gravitational decoupling of the equations of motion, in order to suppress the singularities present in the bulk.

    \item \label{CondicionA} In order to eliminate the singularity located at infinity in the extra dimension, the previous condition must be satisfied, see equations \eqref{Ricci5D} and \eqref{Kre5D}, and additionally, looking the same equations, both the function \(A(y)\) and its first and second derivatives, \(A'(y)\) and \(A''(y)\), must have finite values at every location included the infinity of the extra coordinate $y$.
\end{enumerate}

Some remarks regarding the equations of motion:

\begin{itemize}
    \item The regularity conditions described in point \ref{CondicionRho}, which are related to the conditions that \( m \) and \( \bar{\rho} \) must satisfy (outlined in Appendix \ref{RBH4D}), along with the regularity conditions \ref{CondicionA} for \( {A} \), ensure that the equations of motion of order \( \alpha \) are free from singularities everywhere.
    \item The regularity conditions mentioned in \ref{CondicionA} for \( A' \) and \( A'' \) ensure that the right-hand side of the equations of motion of order \(\alpha^0\) is free of singularities everywhere. This condition also implies that the left-hand side of these equations must be free of singularities.
\end{itemize}

\subsection{Proposed recipe for solving our decoupled equations} \label{Receta}

Our gravitational decoupling background can be interpreted as follows: we have a system $\mathcal{A}$ of order \( \alpha^0 \), which is associated with a seed spacetime and is coupled to the scalar field and potential matter sources. Conversely, the MGD of a seed spacetime gives rise to a system $\mathcal{B}$, governed by the so-called Quasi-Einstein equations, which are of order \( \alpha \) and are coupled to the bulk energy-momentum tensor fields. Therefore, we propose the following procedure for solving the system:

\begin{enumerate}
    \item Solve the \( y \)-dependent equations of motion of order \(\alpha^0\):
    \begin{enumerate}
        \item Choose as a  test of proof a function \( A(y) \) that satisfies all the criteria outlined in point \ref{CondicionA}.
        \item Compute the values of $\phi (y)$ and $V\left (\phi (y) \right)$.
    \end{enumerate}
    \item Solve the $(r,y)$-dependent equations of motion of order \(\alpha^1\):
    \begin{enumerate}
        \item Pick up an already existing solution \( m(r) \) from the literature that satisfies condition \ref{CondicionRho}, which is associated with the values of \( \bar{\rho}, \bar{p}_r, \) and \( \bar{p}_t \). Some examples in references \cite{Hayward:2005gi,Dymnikova:1992ux,Spallucci:2017aod}.
        \item Substitute the values of \( \bar{\rho}, \bar{p}_r, \bar{p}_t \), and \( A(y) \) into the energy-momentum tensor corresponding to the bulk \ref{EMBulkSplit}.
    \end{enumerate}
\end{enumerate}

\subsection{An example of solving the equations of motion}

\subsubsection{\bf Solve the \( y \)-dependent equations of motion of order \(\alpha^0\) } 

As  test of proof we propose the function:

\begin{equation} \label{NuestroWF}
    A(b,y)=- k \sqrt{\frac{y^2+b^2}{b^2y^2+c^2}}
\end{equation}
It is direct to check that: 
\begin{equation} \label{CondicionARS}
        \displaystyle \lim_{b \to 0} A(y,b) = -k(b/c) |y| \sim \mbox{\,RS warp factor,}
    \end{equation}

\begin{align} \label{CondicionAInfinito}
    &\displaystyle \lim_{y \to \pm \infty} A(y,b) = -(k/b) \Rightarrow \mbox{ (finite)} \\
    &\displaystyle \lim_{y \to \pm \infty} {A}'(y,b) =0 \\
    &\displaystyle \lim_{y \to \pm \infty} {A''}(y,b) = 0
\end{align}
here $'$ indicates derivation respect to $y$. In order that $-A(b,y)$ be an increasing function, and thus, $-A(b,y \to \infty)>-A(b,0)$ we impose that 
\begin{equation}
  c>b^2  
\end{equation}

\begin{figure}[h]   
    \centering 
    \includegraphics[scale=.7]{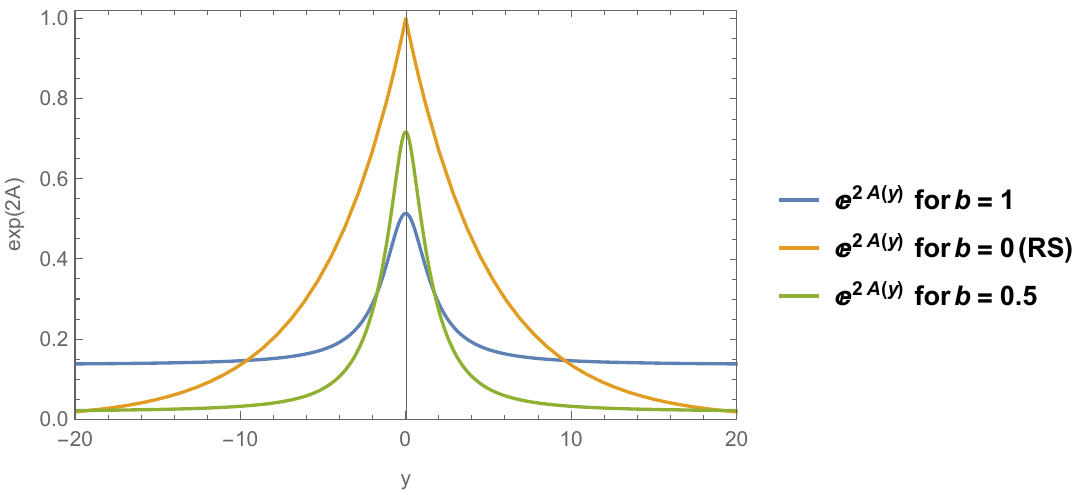} 
    \caption{Warp factor} \label{FigWF}
 \end{figure}

A typical behavior of the warp factor is shown in Figure \ref{FigWF}. It is regular, exhibits $\mathcal{Z}_2$ symmetry, has a peak at the symmetry axis \( y=0 \), and decreases as one moves away from it. In the limit \( b \to 0 \), its behavior resembles that of a thin RS brane. It is worth nothing that, for thick branes, the variation around the peak is smooth.

We can also verify that:
\begin{align} \label{CondicionACero}
    &\displaystyle \lim_{y \to 0} A(y,b) = -k(b/c) \Rightarrow \mbox{ (finite)} \\
    &\displaystyle \lim_{y \to 0} {A}'(y,b) =0 \label{PrimeraDerivada} \\
    &\displaystyle \lim_{y \to 0} {A''}(y,b) = \frac{b^2 (b^2 - c)(b^2 + c) k}{c^6 \left( \frac{b^2}{c^2} \right)^{3/2}} \Rightarrow \mbox{ (finite and negative)} \label{SegundaDerivada}
\end{align}

From equation \eqref{PrimeraDerivada}, we can see that the first derivative of \( A \) is zero, indicating a smoother variation at \( y = 0 \) compared to the RS model, which behaves like a sawtooth at this point. On the other hand, the second derivative, equation \eqref{SegundaDerivada}, has a finite negative value, unlike the RS model, where it behaves like a Dirac delta, blowing up to negative infinity at this point. From this point forward, for simplicity, we will set \( k = 1 \).

Subtracting equations \eqref{ComponenteTAzul} and \eqref{ComponenteYAzul} 

\begin{equation} \label{Phi}
    (\phi')^2=-12M^{3}A''= -12M^{3} \frac{b^2 \left(-c^4 - 2 b^6 y^2 + 2 b^2 c^2 y^2 + 3 c^2 y^4 + b^4 \left(c^2 - 3 y^4\right)\right)}{\left(\frac{b^2 + y^2}{c^2 + b^2 y^2}\right)^{3/2} \left(c^2 + b^2 y^2\right)^4}
\end{equation}

\begin{figure}[h]   
    \centering 
    \includegraphics[scale=.7]{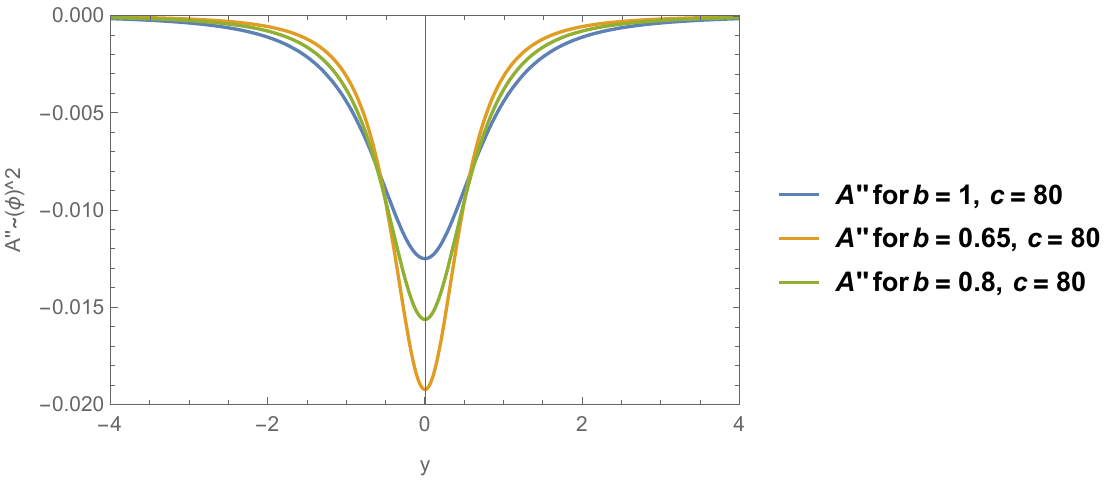} 
    \caption{$A''$} \label{FigDeltaDirac}
 \end{figure}

In relation to the previous paragraph, we can observe in Figure \ref{FigDeltaDirac} that the second derivative of our warp factor has a finite value in the vicinity of \( y = 0 \), which results in a smooth function rather than a Dirac delta distribution. We can also observe that, as $b \to 0$, where the warp factor approaches the RS behavior, the behavior of $A''$ resembles that of a Dirac delta function. Thus,  as we sill see in equation \eqref{Phi} \( (\phi')^2 \sim - A''  \neq -\delta(y) \).

\begin{figure}[h]   
    \centering 
    \includegraphics[scale=.7]{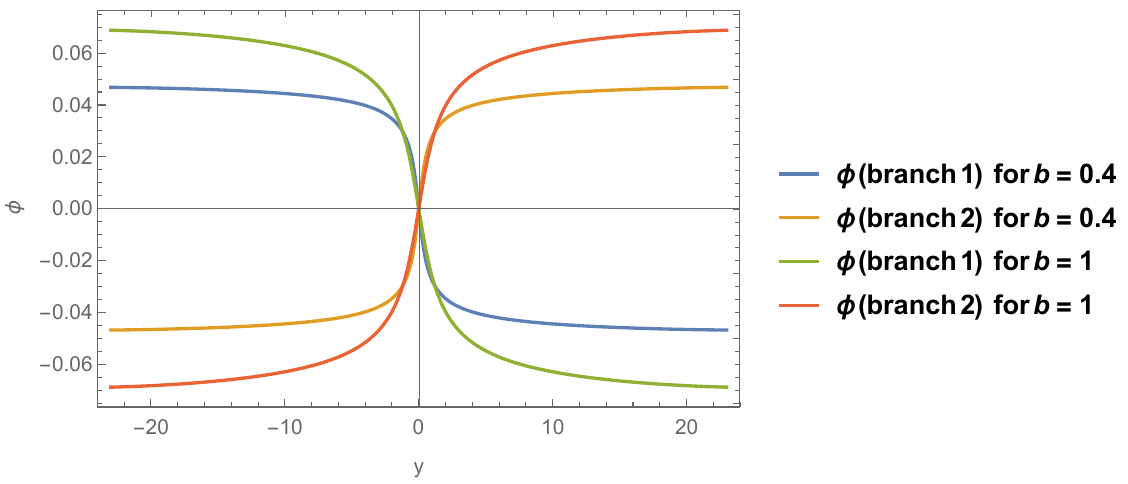} 
    \caption{$\phi(y)$} \label{FigScalarField}
 \end{figure}

Unfortunately, this equation does not have an analytical solution. However, both branches of the solution can be observed numerically. As shown in Figure \ref{FigScalarField}, the behavior is analogous to that of a kink scalar field \cite{Rosa:2022fhl}. The vertical distance between the two asymptotes is comparable to the distance between the two vacua \( (-v \approx \phi_{min}, v \approx \phi_{max}) \) of a kink. Thus, The field \( \phi \) interpolates between $(-v,v)$ as it varies along the extra dimension.

Adding equations \eqref{ComponenteTAzul} and \eqref{ComponenteYAzul} 
\begin{equation}
    V=-6M^3 \left( 4 (A')^2+A''  \right) = -\frac{6 \left( \frac{4 (b^4 - c^2)^2 y^2 (c^2 + b^2 y^2)}{b^2 + y^2} + 
    \frac{b^2 \left(-c^4 - 2 b^6 y^2 + 2 b^2 c^2 y^2 + 3 c^2 y^4 + 
    b^4 (c^2 - 3 y^4)\right)}{\left( \frac{b^2 + y^2}{c^2 + b^2 y^2} \right)^{3/2}} \right)}{(c^2 + b^2 y^2)^4}
\end{equation}

\begin{figure}[h]   
    \centering 
    \includegraphics[scale=.6]{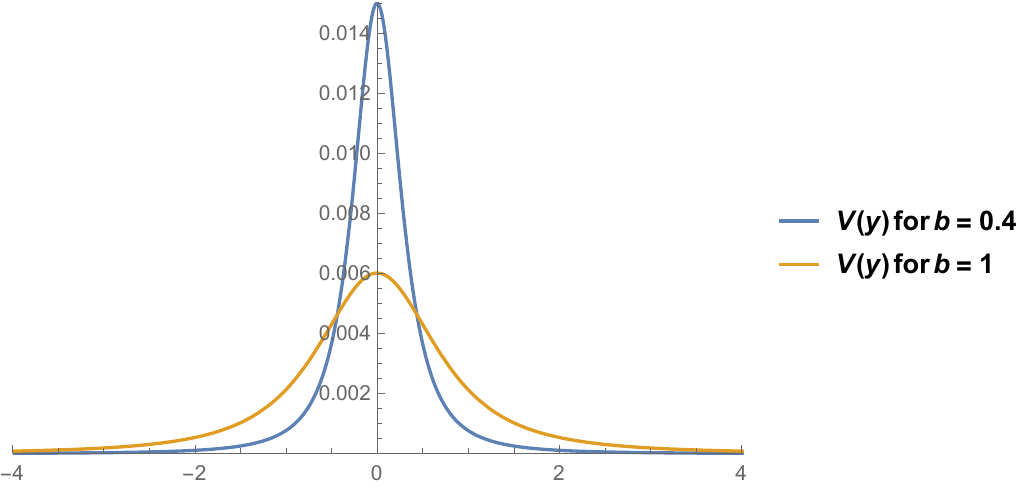} 
    \includegraphics[scale=.6]{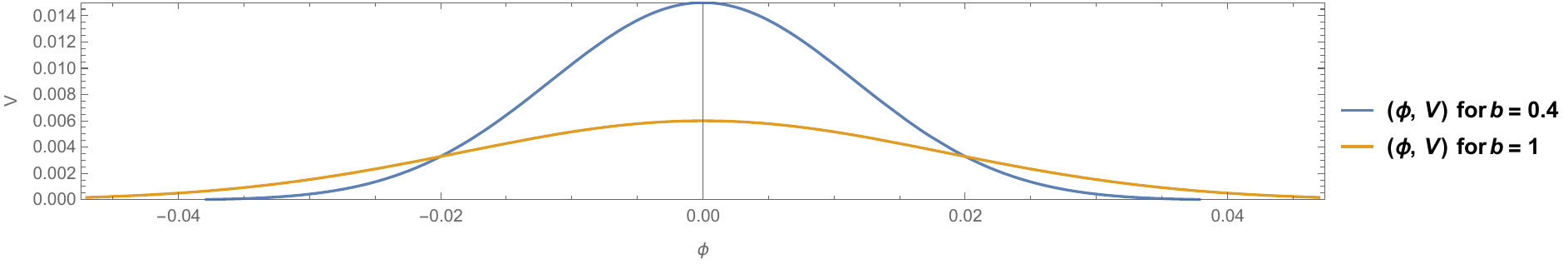} 
    \caption{First: $V(y)$. Second $\phi$ vs $V$} \label{FigPotenciales}
 \end{figure}

In Figure \ref{FigPotenciales}, we observe the behavior of \( V(y) \) and \( \phi \) versus \( V \), respectively. It is evident that both the scalar field and the potential, along with their respective derivatives, have finite values around \( y = 0 \), indicating that the scalar field is smooth. This contrasts with the matter sources featuring a Dirac delta in the Randall-Sundrum model.

Due to the \(\mathcal{Z}_2\) symmetry of our warp factor, we can observe from the graphs that the potential also exhibits this symmetry. There are two degenerate minima of \( V(\phi) \) that occur at \( \phi = \pm v \), which represent the preferred values for the ground state, or so-called vacuum state, of \( \phi \). The scalar field \( \phi \) could be \(-v\) in one particular region of spacetime and \(+v\) in another. This latter scenario corresponds to a domain-wall configuration, where the domain wall is the relatively small region that separates the \(-v\) vacuum from the \(+v\) vacuum.

\paragraph{\bf About the brane limit for our warp factor}: 

It is straightforward to check that, in the limit as \( b \to 0 \), the second and third terms of the action \eqref{AccionInicialSuave} behave as follows:

\begin{align}\label{align}
    \bar{S} \in S  =&- \lim_{b \to 0} \int d^{4}x\int dy\ \sqrt{-g} \biggl(  \dfrac{1}{2}(\partial\phi)^{2} +V(\phi) \biggl) \nonumber \\
                   =& \lim_{b \to 0} 12 M^3 \int \sqrt{-g^{(4D)}} d^{4}x\int dy\ \exp\left (4A(y) \right) \left( 2 (A')^2 + A''   \right) \nonumber \\
                    \sim& 24 M^3 \int \sqrt{-g^{(4D)}} d^{4}x\int dy\ \exp\left (-4k|y| \right) + 12 M^3 \int \sqrt{-g^{(4D)}} d^{4}x\int dy\ \exp\left (4A(y) \right) \delta(y) \nonumber \\
                    \sim& 24 M^3 \int \sqrt{-g} d^{5}x \Lambda_{eff}  + 12 M^3 \sigma \int d^{4}x \sqrt{-g^{(4D)}}
\end{align}
where the first term correspond to the bulk term \cite{Kehagias:2000au} and the second term correspond to the brane term. This, we recovered the RS set--up in the limit $b\to 0$. Thus, \( b \) can be interpreted as a control parameter, such that in the limit \( b \to 0 \), the RS structure is recovered.

\subsubsection{\bf About the solution of the $(r,y)$-dependent Quasi Einstein equations of order \(\alpha^1\):}

As previously mentioned, the novelty of this work does not lie in finding a 4D RBH geometry, but rather in providing a methodology to find a warp factor \( A(y) \) that suppresses the singularity present at infinity in the extra coordinate. Thus, by following step 2a) of the recipe in subsection \ref{Receta}, Any chosen geometry \( m, \bar{\rho}, \bar{p}_r, \bar{p}_t \), when substituted into the energy-momentum tensor (equation \ref{EMBulkSplit}), along with the warp factor \( A(y) \) selected in the previous section, will correspond to a solution of the Quasi-Einstein equations \eqref{ComponenteThetaNegro}, \eqref{ComponenteYNegro}, \eqref{ConservacionYCuasi} and \eqref{ConservacionRCuasi}

\subsubsection{\bf Ricci Scalar and regularity of the geometry :}

The value of the Ricci scalar is given by:

\begin{equation}
R_{5D} (r,y)=   \bar{R}_{4D}(r) \exp\left(2 \sqrt{\frac{b^2 + y^2}{c^2 + b^2 y^2}}\right) + \frac{4 \left( -\frac{5 (b^4 - c^2)^2 y^2 (c^2 + b^2 y^2)}{b^2 + y^2} - \frac{2 b^2 (-c^4 - 2 b^6 y^2 + 2 b^2 c^2 y^2 + 3 c^2 y^4 + b^4 (c^2 - 3 y^4))}{\left( \frac{b^2 + y^2}{c^2 + b^2 y^2} \right)^{3/2}} \right)}{(c^2 + b^2 y^2)^4}
\end{equation}

It is easy to check that the asymptotic value as \( y \to \pm \infty \) is given by:
\begin{equation}
    \displaystyle \lim_{y \to \pm \infty} R_{5D} (r,y) = \bar{R}_{4D}(r) \exp (2/b)
\end{equation}
In this way, the sign of the Ricci curvature invariant in this asymptotic limit depends exclusively on the sign of \( \bar{R}_{4D} \) for a given value of the radial coordinate.

Thus, following the earlier proposal, \( R_{5D} \) is finite everywhere in the bulk, including the limits \( r \to 0 \) and \( y \to \pm \infty \). Similarly, it is direct to  check the same for the five-dimensional Kretschmann scalar given by Equation \eqref{Kre5D}, allowing us to assert that our five-dimensional geometry is regular everywhere.

As mentioned in Appendix \ref{RBH4D}, \( \bar{R}_{4D} (r \approx 0) > 0 \) in the 4D de Sitter core. Then, \( \bar{R}_{4D} (r > r_*) < 0 \), and at the radial asymptote, we have \( \displaystyle \lim_{r \to \infty} \bar{R}_{4D} \to 0 \). Thus, the asymptotic value in the extra coordinate of \( \displaystyle \lim_{y \to \pm \infty} R_{5D} (r,y) \) is

\[
\displaystyle \lim_{y \to \pm \infty} R_{5D} (r,y) \sim 
\begin{cases} 
\text{Positive close to the radial origin at the dS core} \\ 
\text{Negative for } r > r_* \\ 
0 \text{ for the radial asymptote} 
\end{cases}
\]

\begin{figure}[h]   
    \centering 
    \includegraphics[scale=.65]{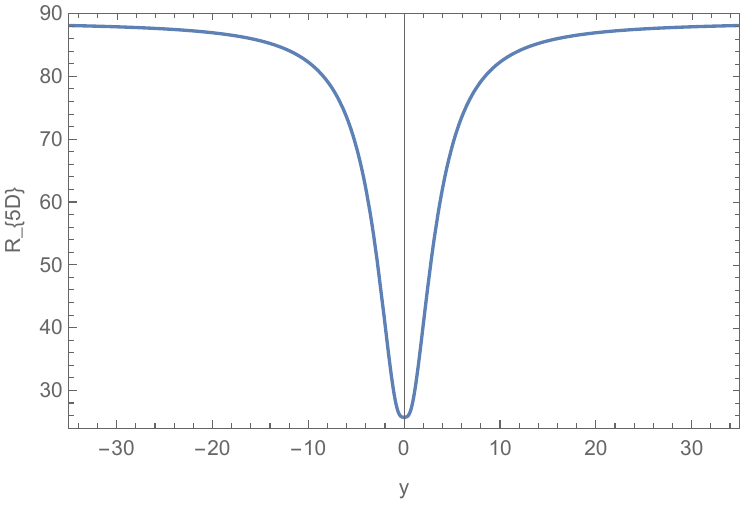} 
    \includegraphics[scale=.65]{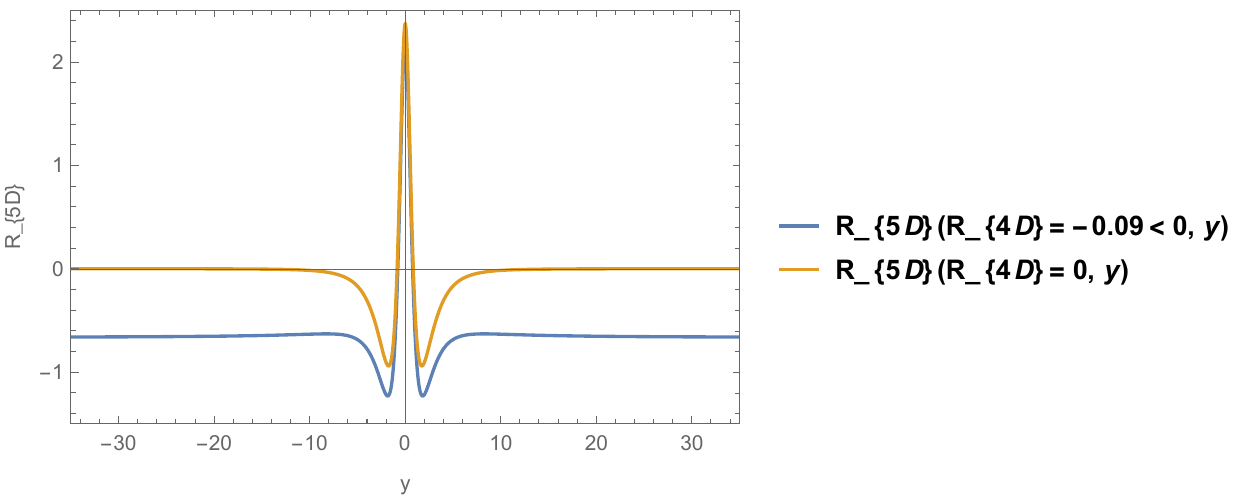} 
    \caption{Ricci scalar $5D$ behavior using $R_{4D}$ oh Hayward model. First close to the radial origin at dS core. Second: for finite $r>r_*$ where $R_{4D}<0$ and for $r\to \infty$ where $R_{4D}\to 0$.} \label{FigRicci5D}
 \end{figure} 

 The generic behavior of the 5D Ricci scalar as a function of the extra coordinate \( y \) is illustrated in Figure \eqref{FigRicci5D} for different fixed locations of the radial coordinate. First, we observe that close to the radial origin at the dS core, \( R_{5D} \) is always positive and approaches a positive asymptote. Secondly, for finite values of \( r > r_* \), where \( R_{4D} < 0 \), the value of \( R_{5D} \) is positive in the vicinity of the symmetry axis \( y = 0 \), then changes to negative at a finite value of \( y \), reaching a minimum, and subsequently increases to approach a negative asymptote. Finally, at the boundary of the radial coordinate, as \( r \to \infty \), where \( R_{4D} \to 0 \), the value of \( R_{5D} \) is positive near the symmetry axis \( y = 0 \), then changes to negative at a finite value of \( y \), reaching a minimum, and subsequently increases to approach an asymptote of zero.

 On the other hand, although the stability under wavelength perturbations is beyond the scope of this work, the particular behavior of the Ricci curvature invariant, which in our case depends on both the radial coordinate and the extra dimension, raises the question of how our geometry would behave under such perturbations. In this context, regarding the specific case where \( \displaystyle \lim_{y \to \pm \infty} R_{5D} \to 0 \), references \cite{Gregory:1993vy,Gregory:2000gf} assert that black hole solutions embedded in an extra dimension within a braneworld framework (the so-called black string) can be unstable to wavelength perturbations in asymptotically flat space. This type of asymptote is also studied in \cite{German:2015cna} to address the stability problem associated with scalar perturbations of tachyonic thick branes. Other scenarios where \( \displaystyle \lim_{y \to \pm \infty} R_{5D} \to 0 \) are discussed in \cite{Koley:2005nv}. Thus, the stability of this type of asymptote, as well as the others previously described, could be investigated in future work.

\subsection{{A Brief Discussion on the Extension of the Horizon in the Bulk}}

{We can easily detect the existence of an event horizon since the induced metric for a fixed value of $y$ in Equation \eqref{NuestroElementoDeLinea} precisely represents a four-dimensional regular black hole. However, a natural question that arises when studying braneworld black holes is whether the event horizon of the black hole geometry extends into the bulk. This problem has been addressed in different contexts for the case of thin branes confined by a Dirac delta. The presence of the latter leads to a specific structure of the induced Einstein equations on the brane \cite{Shiromizu:1999wj}. Using this fact, in reference \cite{Casadio:2002uv}, a method is proposed to determine the shape of the event horizon extension in the bulk. This method is applied for the case where, using the mentioned Dirac delta in the induced equations, the projected Ricci scalar on the brane vanishes, $R^{4D} = 0$, and where the bulk equations in the directions of the brane are such that $R^{5D}_{ij} = \Lambda g_{ij}$ with $i,j = 0,1,2,3$. Astrophysical data of the order of the solar mass were used. The analysis suggests that the horizon does close in the bulk, and a value for the extra coordinate $y = y_0$ is obtained, which marks the extent to which the horizon would propagate in the form of a pancake. In this same framework, reference \cite{Chamblin:2000ra} argues that, in a braneworld black hole where the induced metric corresponds to a charged black hole, the event horizon extends into the dimension transversal to the brane depending on the value of the electric charge parameter.}

{However, it seems that the methodologies described in the references from the previous paragraph should not be applicable to our framework, due to the presence of a scalar field instead of a Dirac delta and the presence of matter fields in the bulk. Thus, a more in-depth study is needed in future work to explore the possible extension of the event horizon for thick branes in the presence of a scalar field.}

\subsubsection{ {\bf Event horizon in the representation of an exponentially localized five-dimensional brane-world black hole.}}

{ Recent work in references \cite{Nakas:2020sey,Nakas:2021srr,Nakas:2023yhj} developed an algorithm that allows for the representation of the geometry of an analytic, exponentially localized five-dimensional brane-world black hole. This approach contrasts with the usual $4D$ black hole embedded in an extra dimension. Although the two representations differ, remarkably, this algorithm enables testing of how the event horizon extends in the bulk under this different representation. Therefore, although our previously described background is devoted to the usual thick brane embedded in an extra dimension, the use of this algorithm, based on the representation of an exponentially localized five-dimensional brane-world black hole, will allow us to test whether it is possible to investigate if the event horizon of out thick brane extends in the bulk under this representation. Thus, although the extension of the event horizon in this representation does not suggest that the event horizon would also extend in the usual background previously described, in this section we present an example of a thick brane-world black hole configuration where the extension of the horizon in the bulk is tested.}

\begin{figure}[h]   
    \centering 
    \includegraphics[scale=.8]{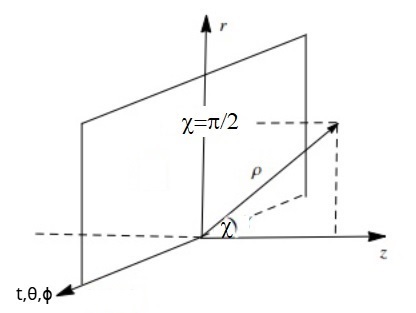} 
    \caption{brane world set up} \label{FigBrana}
 \end{figure}

\begin{figure}[h]   
    \centering 
    \includegraphics[scale=.8]{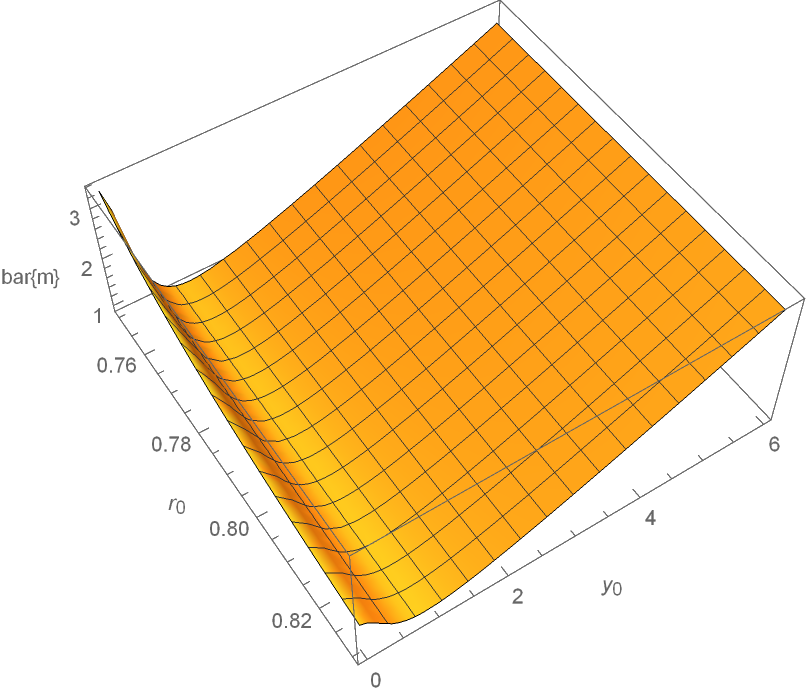}
    \includegraphics[scale=.8]{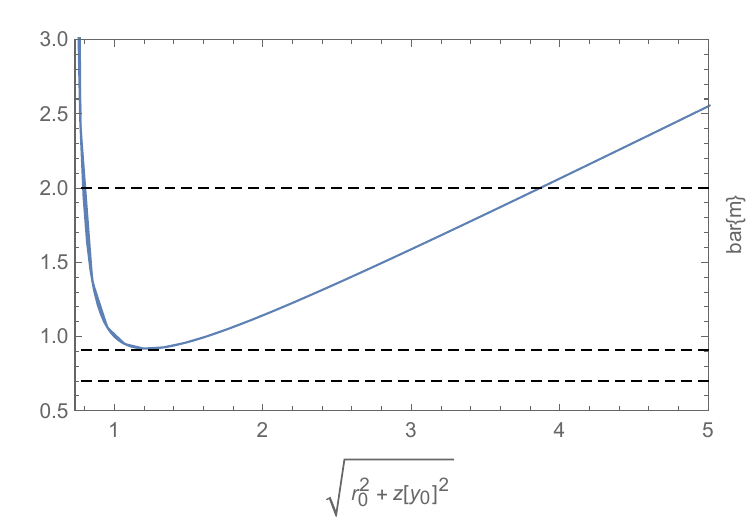} 
    \caption{$\bar{m}$ parameter for $k=1,b=0.4,c=100$} \label{FigParametro}
 \end{figure}

{\begin{enumerate}
    \item Consider the typical line element of RS:
    \begin{equation} \label{TipicoRS}
ds^2 = e^{2A(y)} \left( -dt^2 + dr^2 + r^2 \, d\Omega_2^2 \right) + dy^2
\end{equation}
\item Apply the invertible coordinate transformation $dy^2 = e^{2A(y)} \, dz^2$
and we are led to 
\begin{equation} \label{EcuacionDeZ}
z(y) \equiv \int_0^y dw \, e^{-A(w)} 
\end{equation}
In our case the function $A(y)$ is given by the equation \eqref{NuestroWF}. Thus, we obtain:  
\begin{equation} \label{RSconZ}
ds^2 = e^{2A(y(z))} \left( -dt^2 + dr^2 + r^2 \, d\Omega_2^2 \right) + dz^2
\end{equation}
where $d\Omega_2^2 = d\theta^2 + \sin^2\theta \, d\phi^2$.
\item Next, we rewrite the flat component of the line element \eqref{TipicoRS} in five-dimensional spherical coordinates by using the following transformation:  
\begin{equation}
\{r, z\} \to \{\rho \sin \chi, \, \rho \cos \chi\}, \quad \chi \in [0, \pi]
\end{equation}
where, the inverse transformation is given by:  \begin{equation} \label{TransformacionInv}
\{\rho, \chi\} \to \left\{ \sqrt{r^2 + z^2}, \, \tan^{-1} \left( \frac{r}{z} \right) \right\}
\end{equation}
Thus, the equation \eqref{RSconZ} is:
\begin{equation}
ds^2 = e^{2A(y(\rho \cos \chi))} 
\left( - \, dt^2 + d\rho^2 + \rho^2 \, d\Omega_3^2 \right)
\end{equation}
where $d\Omega_3^2 = d\chi^2 + \sin^2\chi \, d\theta^2 + \sin^2\chi \sin^2\theta \, d\phi^2$.
\item It is proposed to rewrite the last equation as follows:
\begin{equation} \label{ElementoLineaNakas}
ds^2 = e^{2A(y(\rho \cos \chi))} 
\left( -  U(\rho)dt^2 + \frac{d\rho^2}{U(\rho)} + \rho^2 \, d\Omega_3^2 \right)
\end{equation}
It is worth mentioning that the above line element represents an exponentially localized five-dimensional brane-world black hole. See figure \ref{FigBrana}.
\item It is possible to express the line element \eqref{ElementoLineaNakas} in the original coordinate system $\{t, r, \theta, \phi, y\}$ by following the inverse transformations \eqref{TransformacionInv}:
\begin{align}
& ds^2 = e^{2A(y)} \Big[ -U(r, y) \, dt^2 + \frac{dr^2}{r^2 + z^2(y)} \left( r^2/U(r, y) + z^2(y) \right) + r^2 d\Omega^2_2 \nonumber \\
& + \frac{2rz(y) e^{A(y)}}{r^2 + z^2(y)} \left ( 1/U(r, y)-1 \right ) \, dr \, dy \Big] \nonumber \\
& + \frac{dy^2}{r^2 + z^2(y)} \left( r^2 + z^2(y)/ U(r, y) \right) 
\end{align}
We can see that the induced metric at $y,z \to 0$ is
\begin{equation}
ds^2= e^{2A(0)} \left( -U(r)dt^2+U(r)^{-1}dr^2+r^2 d\Omega_2^2 \right)
\end{equation}
where $e^{2A(0)}$ is given by the equation \eqref{CondicionACero}. 
\item In reference \cite{Nakas:2020sey} is defined $U(r, y_0) =1-\bar{m}/\sqrt{r^2+z(y_0)^2}$. As a test for this subsection, we will use a generalization of Hayward's mass function (since Dymnikova's mass function grows exponentially and very rapidly). In order to be consistent with the algorithm, we will define the $U$ function as:
\begin{equation}
    U(r,y_0)=1-\alpha \cdot \frac{m(r,y_0)}{\sqrt{r^2+z(y_0)^2}}=1-\alpha \cdot \bar{m} \frac{(\sqrt{r^2+z(y_0)^2}))^3}{\left((\sqrt{r^2+z(y_0)^2})^3+\bar{m} \right) \sqrt{r^2+z(y_0)^2}}
\end{equation}
where $\bar{m}$ is a constant and $z(y_0)$ is given by equation \eqref{EcuacionDeZ} for $y=y_0$. We define the mass parameter to the value of $\bar{m}$ in the parameter space such that $U(r_0, y_0) = 0$. 
\begin{equation}
    \bar{m}= \frac{(r_0^2+z(y_0)^2)\sqrt{r_0^2+z(y_0)^2}}{-1+2r_0^2+2z(y_0)^2}
\end{equation}
Thus, for a value of $\bar{m}$, there exist one or more ordered pairs $\left( \bar{m}, (r_0, y_0) \right)$ such that $U(r_0, y_0) = 0$. The smallest value of $r_0$ represents the inner horizon, and the largest value represents the event horizon. Since $z(y)$ cannot be computed analytically, in Figures \ref{FigParametro}, we compute the value of $\bar{m}$ numerically We can intuitively notice in the figure above that there are two ordered pairs $(r_0 = r_i, y_0 = y_i)$ and $(r_0 = r_h, y_0 = y_h)$ such that $r_i < r_h$ and $y_i < y_h$, and both ordered pairs correspond to the same value of the mass parameter $\bar{m}$. In this way, we can intuit that the inner horizon would be located at $(r_0 = r_i, y_0 = y_i)$ and the event horizon would be located at $(r_0 = r_h, y_0 = y_h)$. Furthermore, in the graph below, $\sqrt{r_0^2 + z(y_0)^2}$ is plotted on the horizontal axis and $\bar{m}$ on the vertical axis. From this graph, we can infer that there is a critical value $\bar{m} = \bar{m}_c$ such that for $\bar{m} < \bar{m}_c$ there is no black hole, for $\bar{m} = \bar{m}_c$ we have an extremal black hole, and for $\bar{m} > \bar{m}_c$, we have a black hole with an inner horizon (located on the curve with negative slope) and an event horizon (located on the curve with positive slope). Thus, we can intuit that both the inner horizon $r_i$ and the event horizon $r_h$ may extend into the bulk for values $y_i$ and $y_h$, respectively, in the parameter space. Unfortunately, since our analysis is numerical, it is difficult to estimate whether the horizon extends in the bulk in the form of a pancake as in \cite{Nakas:2020sey} or another shape, and whether there is a value in the bulk at which the event horizon ceases to exist It is worth mentioning that the representation of an exponentially localized five-dimensional brane-world black hole also implies the variation of other physical features, such as energy conditions, thermodynamics, etc. These aspects go beyond the scope of this work, as in this paper, we have only provided an overview of the potential extension of the event horizon in the bulk.
\end{enumerate}}

\section{Discussion and Summary}

In this work, we have presented a methodology based on the gravitational decoupling of the bulk equations of motion to eliminate the singularity at infinity in the extra coordinate, which typically arises due to the embedding of a 4D black hole in a braneworld setup.

To establish our methodology, we use the Minkowski spacetime embedded in an extra dimension as the $4D$ seed geometry, to which a Minimal Geometric Deformation (MGD) is applied. Thus, the gravitational system decouples in the following manner: a system $\mathcal{A}$ of order $ \alpha^0 $, which is coupled to the scalar field and potential matter sources. Conversely, the MGD of Minkowski space leads to the system $\mathcal{B}$, related with the so-called Quasi-Einstein equations, which are of order \( \alpha \) and are coupled to the bulk energy-momentum tensor fields. Additionally, we have determined the properties that the bulk energy-momentum tensor must satisfy. It is important to note that both gravitational sectors $\mathcal{A}$ and $\mathcal{B}$ do not exchange energy and momentum; rather, there is only gravitational interaction between them. We will outline the constraints that must be satisfied to eliminate the singularity present both at the origin of the radial coordinate and at infinity in the extra coordinate. Additionally, we will provide a recipe for solving our equations of motion. Consequently, the equations of order \( \alpha \) eliminate the singularity at the radial origin, while the equations of order \( \alpha^0 \) eliminate the singularity at infinity in the extra dimension.

We obtain a regular geometry whose warp factor exhibits \(\mathcal{Z}_2\) symmetry, has a peak at the symmetry axis \( y=0 \), and decreases as one moves away from it. In the limit \( b \to 0 \), its behavior resembles that of a thin RS brane. It is worth noting that, for thick branes, the variation around the peak is smooth. The first derivative of the warp factor is zero at the symmetry axis \( y=0 \), indicating a smoother variation at this point compared to the RS model, which behaves like a sawtooth function. On the other hand, the second derivative of the warp factor has a finite negative value, in contrast to the RS model, where it behaves like a Dirac delta function, blowing up to negative infinity at this point. Thus, the pathology of the equations of motion associated with the presence of the Dirac delta function is alleviated. We can also observe that as \( b \to 0 \), where the warp factor approaches RS behavior, the behavior of \( A'' \) resembles that of a Dirac delta function.

We have numerically plotted the behavior of the scalar field in Figure \ref{FigScalarField}, which possesses two branches, resembling the behavior of a kink scalar field \cite{Rosa:2022fhl}. The vertical distance between the two asymptotes is comparable to the distance between the two vacua \( (-v \approx \phi_{\text{min}}, v \approx \phi_{\text{max}}) \) of a kink. Thus, the field \( \phi \) interpolates between \( (-v, v) \) as it varies along the extra dimension. Both the scalar field and the potential, along with their respective derivatives, have finite values around \( y = 0 \), indicating that the scalar field is smooth. This contrasts with the matter sources featuring a Dirac delta function in the Randall-Sundrum model. Due to the \(\mathcal{Z}_2\) symmetry, the potential has two degenerate minima, which occur at \( \phi = \pm v \) and represent the preferred values for the ground state, or vacuum state, of \( \phi \). The scalar field \( \phi \) could be \(-v\) in one particular region of spacetime and \(+v\) in another. This latter scenario corresponds to a domain-wall configuration, where the domain wall is the relatively small region that separates the \(-v\) vacuum from the \(+v\) vacuum.

We have analyzed the behavior of the curvature \( R_{5D} \) as a function of the extra coordinate \( y \) for fixed values of the radial coordinate. We have determined that the asymptotic value of \( R_{5D} \) at \( y \) depends on the value of \( \bar{R}_{4D}(r) \). Specifically, for values close to the de Sitter core located at the radial origin, the asymptote of \( R_{5D} \) at \( y \) is positive. For values of \( r > r_* \), where \( \bar{R}_{4D}(r) \) is negative, the asymptote of \( R_{5D} \) at \( y \) is negative. At the radial boundary of the geometry, where \( \bar{R}_{4D}(r) \) transitions, the asymptote of \( R_{5D} \) at \( y \) is zero.

Although this work presents a methodology aimed at eliminating the singularity at infinity, it is expected that the provided methodology will serve as a basis for studying other future problems related to the embedding of gravitational objects in a braneworld setup. For more complex scenarios, this would likely require redefining the coupling constants and the matter sources coupled in the bulk.

As mentioned in the introduction, there has been significant interest in recent years in testing braneworld scenarios with experimental data \cite{Banerjee:2021aln, Banerjee:2022jog, Visinelli:2017bny,Mishra:2021waw}. For example, reference \cite{Banerjee:2022jog} demonstrated that, by embedding wormhole geometries in a braneworld scenario, the shadow data and the image diameter of Sgr A* fit better within braneworld models compared to the corresponding situation in General Relativity. Thus, from a theoretical point of view, it would be interesting if our methodology could serve as a basis for studying the embedding of other gravitational objects in a braneworld setup. For instance, one could investigate the gravitational decoupling related to the embedding of traversable wormholes, one-way wormholes, or regular black holes in a Simpson-Visser black bounce scenario \cite{Crispim:2024nou,Crispim:2024yjz}. This would require a suitable redefinition of the radial coordinate and an appropriate dependence of both this coordinate and the function \( m \) on the coupling constants. This could be explored in future work

\section*{Acknowledgements}
Milko Estrada is funded by ANID , FONDECYT de Iniciaci\'on en Investigación 2023, Folio 11230247. Tiago M. Crispim and Geová Alencar would like to thank Conselho Nacional de Desenvolvimento Científico e Tecnológico (CNPq) and Fundação Cearense de Apoio ao Desenvolvimento Científico e Tecnológico (FUNCAP) for the financial support.

\appendix

\section{Characteristics of the $4D$ RBH geometry} \label{RBH4D}

Typically, $4D$ RBH geometries exhibit the following characteristics:
\begin{itemize}
    \item The radial energy density, $\bar{\rho}$, reaches a maximum finite near the origin at $r=0$ and continuously decreases, satisfying $\displaystyle \lim_{r \to \infty} \bar{\rho} \to 0$. Consequently, the four-dimensional solution is asymptotically flat. 
    \item The mass function $m(r)$ behaves near the origin as $m \sim r^3$, and is a continuously increasing function. Since the $4D$ geometry is asymptotically flat, it approaches an asymptotic value, $\displaystyle \lim_{r \to \infty} m(r) = \bar{M}$, where $\bar{M}$ is a constant.
\end{itemize}

The two aforementioned points lead to:

\begin{itemize}
    \item The geometry behaves like a maximally symmetric space near the radial origin.
    \item The geometry behaves like a flat Minkowski space at radial infinity.
\end{itemize}

It is worth noting that for our type of 4D RBH, the behavior of the Ricci scalar \( \bar{R}_{4D} \) is generic, as illustrated in Figure \ref{FigRicci4D}, using the Dymnikova and Hayward models as examples. In this regard, we can highlight the following characteristics, which will be useful for analyzing the curvature of the entire 5D spacetime:

\begin{figure}[h]   
    \centering 
    \includegraphics[scale=.5]{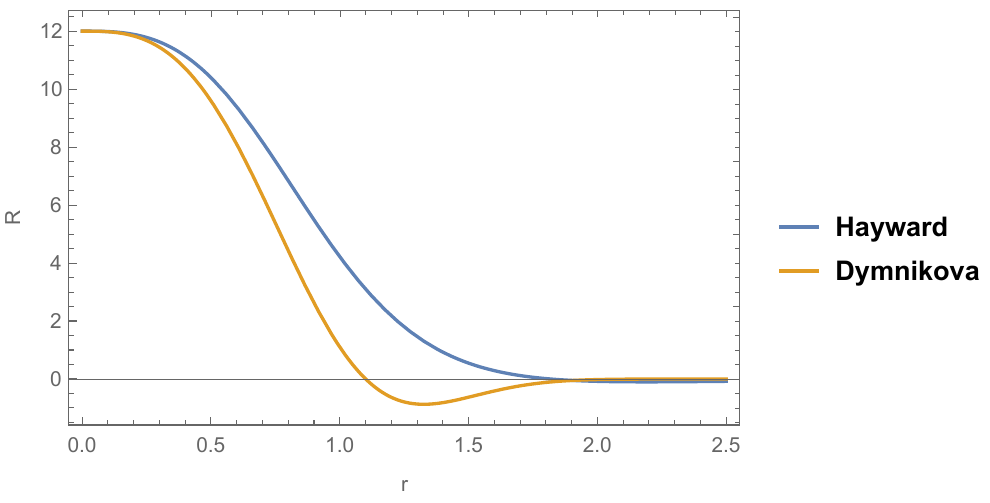} 
    \includegraphics[scale=.5]{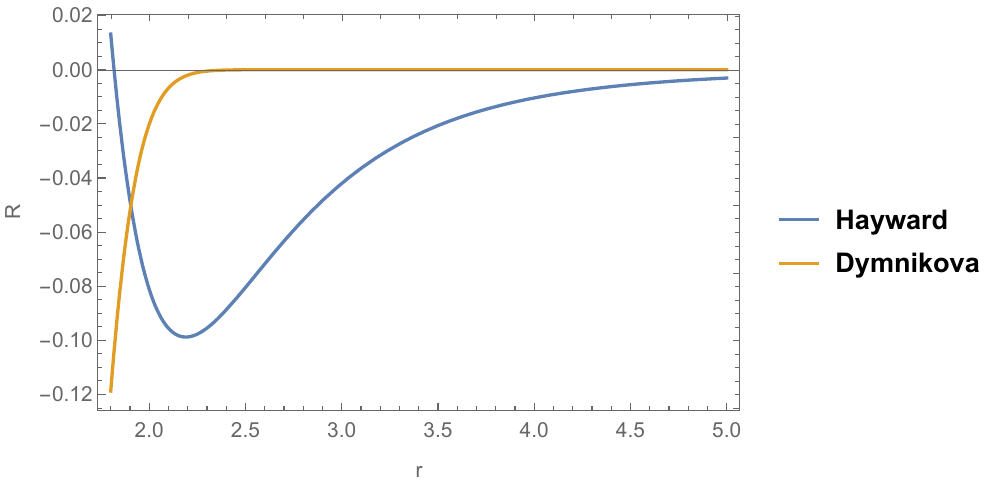} 
    \caption{Ricci scalar $4D$} \label{FigRicci4D}
 \end{figure} 

\begin{itemize}
    \item Near the radial origin, \( r=0 \), in the de Sitter core, \( \bar{R}_{4D} \) has a positive maximum value.
    \item From \( r=0 \), \( \bar{R}_{4D} \) decreases, changing sign to negative at a point we will refer to as \( r=r_* \), until it reaches a minimum value.
    \item From this negative minimum, \( \bar{R}_{4D} \) is a function that approaches \( \bar{R}_{4D} \approx 0 \) asymptotically.
\end{itemize}
 
\bibliography{mybib1}

\begin{thebibliography}{53}%
\makeatletter
\providecommand \@ifxundefined [1]{%
 \@ifx{#1\undefined}
}%
\providecommand \@ifnum [1]{%
 \ifnum #1\expandafter \@firstoftwo
 \else \expandafter \@secondoftwo
 \fi
}%
\providecommand \@ifx [1]{%
 \ifx #1\expandafter \@firstoftwo
 \else \expandafter \@secondoftwo
 \fi
}%
\providecommand \natexlab [1]{#1}%
\providecommand \enquote  [1]{``#1''}%
\providecommand \bibnamefont  [1]{#1}%
\providecommand \bibfnamefont [1]{#1}%
\providecommand \citenamefont [1]{#1}%
\providecommand \href@noop [0]{\@secondoftwo}%
\providecommand \href [0]{\begingroup \@sanitize@url \@href}%
\providecommand \@href[1]{\@@startlink{#1}\@@href}%
\providecommand \@@href[1]{\endgroup#1\@@endlink}%
\providecommand \@sanitize@url [0]{\catcode `\\12\catcode `\$12\catcode `\&12\catcode `\#12\catcode `\^12\catcode `\_12\catcode `\%12\relax}%
\providecommand \@@startlink[1]{}%
\providecommand \@@endlink[0]{}%
\providecommand \url  [0]{\begingroup\@sanitize@url \@url }%
\providecommand \@url [1]{\endgroup\@href {#1}{\urlprefix }}%
\providecommand \urlprefix  [0]{URL }%
\providecommand \Eprint [0]{\href }%
\providecommand \doibase [0]{http://dx.doi.org/}%
\providecommand \selectlanguage [0]{\@gobble}%
\providecommand \bibinfo  [0]{\@secondoftwo}%
\providecommand \bibfield  [0]{\@secondoftwo}%
\providecommand \translation [1]{[#1]}%
\providecommand \BibitemOpen [0]{}%
\providecommand \bibitemStop [0]{}%
\providecommand \bibitemNoStop [0]{.\EOS\space}%
\providecommand \EOS [0]{\spacefactor3000\relax}%
\providecommand \BibitemShut  [1]{\csname bibitem#1\endcsname}%
\let\auto@bib@innerbib\@empty
\bibitem [{\citenamefont {Abbott}\ \emph {et~al.}(2016)\citenamefont {Abbott} \emph {et~al.}}]{LIGOScientific:2016aoc}%
  \BibitemOpen
  \bibfield  {author} {\bibinfo {author} {\bibfnamefont {B.~P.}\ \bibnamefont {Abbott}} \emph {et~al.} (\bibinfo {collaboration} {LIGO Scientific, Virgo}),\ }\bibfield  {title} {\enquote {\bibinfo {title} {{Observation of Gravitational Waves from a Binary Black Hole Merger}},}\ }\href {\doibase 10.1103/PhysRevLett.116.061102} {\bibfield  {journal} {\bibinfo  {journal} {Phys. Rev. Lett.}\ }\textbf {\bibinfo {volume} {116}},\ \bibinfo {pages} {061102} (\bibinfo {year} {2016})},\ \Eprint {http://arxiv.org/abs/1602.03837} {arXiv:1602.03837 [gr-qc]} \BibitemShut {NoStop}%
\bibitem [{\citenamefont {Abbott}\ \emph {et~al.}(2017)\citenamefont {Abbott} \emph {et~al.}}]{LIGOScientific:2017vwq}%
  \BibitemOpen
  \bibfield  {author} {\bibinfo {author} {\bibfnamefont {B.~P.}\ \bibnamefont {Abbott}} \emph {et~al.} (\bibinfo {collaboration} {LIGO Scientific, Virgo}),\ }\bibfield  {title} {\enquote {\bibinfo {title} {{GW170817: Observation of Gravitational Waves from a Binary Neutron Star Inspiral}},}\ }\href {\doibase 10.1103/PhysRevLett.119.161101} {\bibfield  {journal} {\bibinfo  {journal} {Phys. Rev. Lett.}\ }\textbf {\bibinfo {volume} {119}},\ \bibinfo {pages} {161101} (\bibinfo {year} {2017})},\ \Eprint {http://arxiv.org/abs/1710.05832} {arXiv:1710.05832 [gr-qc]} \BibitemShut {NoStop}%
\bibitem [{\citenamefont {Akiyama}\ \emph {et~al.}(2022)\citenamefont {Akiyama} \emph {et~al.}}]{EventHorizonTelescope:2022wkp}%
  \BibitemOpen
  \bibfield  {author} {\bibinfo {author} {\bibfnamefont {Kazunori}\ \bibnamefont {Akiyama}} \emph {et~al.} (\bibinfo {collaboration} {Event Horizon Telescope}),\ }\bibfield  {title} {\enquote {\bibinfo {title} {{First Sagittarius A* Event Horizon Telescope Results. I. The Shadow of the Supermassive Black Hole in the Center of the Milky Way}},}\ }\href {\doibase 10.3847/2041-8213/ac6674} {\bibfield  {journal} {\bibinfo  {journal} {Astrophys. J. Lett.}\ }\textbf {\bibinfo {volume} {930}},\ \bibinfo {pages} {L12} (\bibinfo {year} {2022})},\ \Eprint {http://arxiv.org/abs/2311.08680} {arXiv:2311.08680 [astro-ph.HE]} \BibitemShut {NoStop}%
\bibitem [{\citenamefont {Banerjee}\ \emph {et~al.}(2021)\citenamefont {Banerjee}, \citenamefont {Chakraborty},\ and\ \citenamefont {SenGupta}}]{Banerjee:2021aln}%
  \BibitemOpen
  \bibfield  {author} {\bibinfo {author} {\bibfnamefont {Indrani}\ \bibnamefont {Banerjee}}, \bibinfo {author} {\bibfnamefont {Sumanta}\ \bibnamefont {Chakraborty}}, \ and\ \bibinfo {author} {\bibfnamefont {Soumitra}\ \bibnamefont {SenGupta}},\ }\bibfield  {title} {\enquote {\bibinfo {title} {{Looking for extra dimensions in the observed quasi-periodic oscillations of black holes}},}\ }\href {\doibase 10.1088/1475-7516/2021/09/037} {\bibfield  {journal} {\bibinfo  {journal} {JCAP}\ }\textbf {\bibinfo {volume} {09}},\ \bibinfo {pages} {037} (\bibinfo {year} {2021})},\ \Eprint {http://arxiv.org/abs/2105.06636} {arXiv:2105.06636 [gr-qc]} \BibitemShut {NoStop}%
\bibitem [{\citenamefont {Banerjee}\ \emph {et~al.}(2022)\citenamefont {Banerjee}, \citenamefont {Chakraborty},\ and\ \citenamefont {SenGupta}}]{Banerjee:2022jog}%
  \BibitemOpen
  \bibfield  {author} {\bibinfo {author} {\bibfnamefont {Indrani}\ \bibnamefont {Banerjee}}, \bibinfo {author} {\bibfnamefont {Sumanta}\ \bibnamefont {Chakraborty}}, \ and\ \bibinfo {author} {\bibfnamefont {Soumitra}\ \bibnamefont {SenGupta}},\ }\bibfield  {title} {\enquote {\bibinfo {title} {{Hunting extra dimensions in the shadow of Sgr A*}},}\ }\href {\doibase 10.1103/PhysRevD.106.084051} {\bibfield  {journal} {\bibinfo  {journal} {Phys. Rev. D}\ }\textbf {\bibinfo {volume} {106}},\ \bibinfo {pages} {084051} (\bibinfo {year} {2022})},\ \Eprint {http://arxiv.org/abs/2207.09003} {arXiv:2207.09003 [gr-qc]} \BibitemShut {NoStop}%
\bibitem [{\citenamefont {Visinelli}\ \emph {et~al.}(2018)\citenamefont {Visinelli}, \citenamefont {Bolis},\ and\ \citenamefont {Vagnozzi}}]{Visinelli:2017bny}%
  \BibitemOpen
  \bibfield  {author} {\bibinfo {author} {\bibfnamefont {Luca}\ \bibnamefont {Visinelli}}, \bibinfo {author} {\bibfnamefont {Nadia}\ \bibnamefont {Bolis}}, \ and\ \bibinfo {author} {\bibfnamefont {Sunny}\ \bibnamefont {Vagnozzi}},\ }\bibfield  {title} {\enquote {\bibinfo {title} {{Brane-world extra dimensions in light of GW170817}},}\ }\href {\doibase 10.1103/PhysRevD.97.064039} {\bibfield  {journal} {\bibinfo  {journal} {Phys. Rev. D}\ }\textbf {\bibinfo {volume} {97}},\ \bibinfo {pages} {064039} (\bibinfo {year} {2018})},\ \Eprint {http://arxiv.org/abs/1711.06628} {arXiv:1711.06628 [gr-qc]} \BibitemShut {NoStop}%
\bibitem [{\citenamefont {Mishra}\ \emph {et~al.}(2022)\citenamefont {Mishra}, \citenamefont {Ghosh},\ and\ \citenamefont {Chakraborty}}]{Mishra:2021waw}%
  \BibitemOpen
  \bibfield  {author} {\bibinfo {author} {\bibfnamefont {Akash~K.}\ \bibnamefont {Mishra}}, \bibinfo {author} {\bibfnamefont {Abhirup}\ \bibnamefont {Ghosh}}, \ and\ \bibinfo {author} {\bibfnamefont {Sumanta}\ \bibnamefont {Chakraborty}},\ }\bibfield  {title} {\enquote {\bibinfo {title} {{Constraining extra dimensions using observations of black hole quasi-normal modes}},}\ }\href {\doibase 10.1140/epjc/s10052-022-10788-x} {\bibfield  {journal} {\bibinfo  {journal} {Eur. Phys. J. C}\ }\textbf {\bibinfo {volume} {82}},\ \bibinfo {pages} {820} (\bibinfo {year} {2022})},\ \Eprint {http://arxiv.org/abs/2106.05558} {arXiv:2106.05558 [gr-qc]} \BibitemShut {NoStop}%
\bibitem [{\citenamefont {Casadio}\ and\ \citenamefont {Ovalle}(2012)}]{Casadio:2012pu}%
  \BibitemOpen
  \bibfield  {author} {\bibinfo {author} {\bibfnamefont {R.}~\bibnamefont {Casadio}}\ and\ \bibinfo {author} {\bibfnamefont {J.}~\bibnamefont {Ovalle}},\ }\bibfield  {title} {\enquote {\bibinfo {title} {{Brane-world stars and (microscopic) black holes}},}\ }\href {\doibase 10.1016/j.physletb.2012.07.041} {\bibfield  {journal} {\bibinfo  {journal} {Phys. Lett. B}\ }\textbf {\bibinfo {volume} {715}},\ \bibinfo {pages} {251--255} (\bibinfo {year} {2012})},\ \Eprint {http://arxiv.org/abs/1201.6145} {arXiv:1201.6145 [gr-qc]} \BibitemShut {NoStop}%
\bibitem [{\citenamefont {da~Rocha}\ and\ \citenamefont {Coimbra-Araujo}(2006)}]{daRocha:2006ei}%
  \BibitemOpen
  \bibfield  {author} {\bibinfo {author} {\bibfnamefont {Roldao}\ \bibnamefont {da~Rocha}}\ and\ \bibinfo {author} {\bibfnamefont {Carlos~H.}\ \bibnamefont {Coimbra-Araujo}},\ }\bibfield  {title} {\enquote {\bibinfo {title} {{Extra dimensions in LHC via mini-black holes: Effective Kerr-Newman brane-world effects}},}\ }\href {\doibase 10.1103/PhysRevD.74.055006} {\bibfield  {journal} {\bibinfo  {journal} {Phys. Rev. D}\ }\textbf {\bibinfo {volume} {74}},\ \bibinfo {pages} {055006} (\bibinfo {year} {2006})},\ \Eprint {http://arxiv.org/abs/hep-ph/0607027} {arXiv:hep-ph/0607027} \BibitemShut {NoStop}%
\bibitem [{\citenamefont {Das}\ \emph {et~al.}(2020)\citenamefont {Das}, \citenamefont {Kumar},\ and\ \citenamefont {Samanta}}]{Das:2020gie}%
  \BibitemOpen
  \bibfield  {author} {\bibinfo {author} {\bibfnamefont {Goutam}\ \bibnamefont {Das}}, \bibinfo {author} {\bibfnamefont {M.~C.}\ \bibnamefont {Kumar}}, \ and\ \bibinfo {author} {\bibfnamefont {Kajal}\ \bibnamefont {Samanta}},\ }\bibfield  {title} {\enquote {\bibinfo {title} {{Resummed inclusive cross-section in Randall-Sundrum model at NNLO+NNLL}},}\ }\href {\doibase 10.1007/JHEP07(2020)040} {\bibfield  {journal} {\bibinfo  {journal} {JHEP}\ }\textbf {\bibinfo {volume} {07}},\ \bibinfo {pages} {040} (\bibinfo {year} {2020})},\ \Eprint {http://arxiv.org/abs/2004.03938} {arXiv:2004.03938 [hep-ph]} \BibitemShut {NoStop}%
\bibitem [{\citenamefont {Belis}\ \emph {et~al.}(2024)\citenamefont {Belis}, \citenamefont {Wo\'zniak}, \citenamefont {Puljak}, \citenamefont {Barkoutsos}, \citenamefont {Dissertori}, \citenamefont {Grossi}, \citenamefont {Pierini}, \citenamefont {Reiter}, \citenamefont {Tavernelli},\ and\ \citenamefont {Vallecorsa}}]{Belis:2023atb}%
  \BibitemOpen
  \bibfield  {author} {\bibinfo {author} {\bibfnamefont {Vasilis}\ \bibnamefont {Belis}}, \bibinfo {author} {\bibfnamefont {Kinga~Anna}\ \bibnamefont {Wo\'zniak}}, \bibinfo {author} {\bibfnamefont {Ema}\ \bibnamefont {Puljak}}, \bibinfo {author} {\bibfnamefont {Panagiotis}\ \bibnamefont {Barkoutsos}}, \bibinfo {author} {\bibfnamefont {G\"unther}\ \bibnamefont {Dissertori}}, \bibinfo {author} {\bibfnamefont {Michele}\ \bibnamefont {Grossi}}, \bibinfo {author} {\bibfnamefont {Maurizio}\ \bibnamefont {Pierini}}, \bibinfo {author} {\bibfnamefont {Florentin}\ \bibnamefont {Reiter}}, \bibinfo {author} {\bibfnamefont {Ivano}\ \bibnamefont {Tavernelli}}, \ and\ \bibinfo {author} {\bibfnamefont {Sofia}\ \bibnamefont {Vallecorsa}},\ }\bibfield  {title} {\enquote {\bibinfo {title} {{Quantum anomaly detection in the latent space of proton collision events at the LHC}},}\ }\href {\doibase 10.1038/s42005-024-01811-6} {\bibfield  {journal} {\bibinfo  {journal} {Commun. Phys.}\ }\textbf {\bibinfo {volume} {7}},\ \bibinfo
  {pages} {334} (\bibinfo {year} {2024})},\ \Eprint {http://arxiv.org/abs/2301.10780} {arXiv:2301.10780 [quant-ph]} \BibitemShut {NoStop}%
\bibitem [{\citenamefont {Randall}\ and\ \citenamefont {Sundrum}(1999{\natexlab{a}})}]{Randall:1999vf}%
  \BibitemOpen
  \bibfield  {author} {\bibinfo {author} {\bibfnamefont {Lisa}\ \bibnamefont {Randall}}\ and\ \bibinfo {author} {\bibfnamefont {Raman}\ \bibnamefont {Sundrum}},\ }\bibfield  {title} {\enquote {\bibinfo {title} {{An Alternative to compactification}},}\ }\href {\doibase 10.1103/PhysRevLett.83.4690} {\bibfield  {journal} {\bibinfo  {journal} {Phys. Rev. Lett.}\ }\textbf {\bibinfo {volume} {83}},\ \bibinfo {pages} {4690--4693} (\bibinfo {year} {1999}{\natexlab{a}})},\ \Eprint {http://arxiv.org/abs/hep-th/9906064} {arXiv:hep-th/9906064} \BibitemShut {NoStop}%
\bibitem [{\citenamefont {Randall}\ and\ \citenamefont {Sundrum}(1999{\natexlab{b}})}]{Randall:1999ee}%
  \BibitemOpen
  \bibfield  {author} {\bibinfo {author} {\bibfnamefont {Lisa}\ \bibnamefont {Randall}}\ and\ \bibinfo {author} {\bibfnamefont {Raman}\ \bibnamefont {Sundrum}},\ }\bibfield  {title} {\enquote {\bibinfo {title} {{A Large mass hierarchy from a small extra dimension}},}\ }\href {\doibase 10.1103/PhysRevLett.83.3370} {\bibfield  {journal} {\bibinfo  {journal} {Phys. Rev. Lett.}\ }\textbf {\bibinfo {volume} {83}},\ \bibinfo {pages} {3370--3373} (\bibinfo {year} {1999}{\natexlab{b}})},\ \Eprint {http://arxiv.org/abs/hep-ph/9905221} {arXiv:hep-ph/9905221} \BibitemShut {NoStop}%
\bibitem [{\citenamefont {Chamblin}\ \emph {et~al.}(2000)\citenamefont {Chamblin}, \citenamefont {Hawking},\ and\ \citenamefont {Reall}}]{Chamblin:1999by}%
  \BibitemOpen
  \bibfield  {author} {\bibinfo {author} {\bibfnamefont {A.}~\bibnamefont {Chamblin}}, \bibinfo {author} {\bibfnamefont {S.~W.}\ \bibnamefont {Hawking}}, \ and\ \bibinfo {author} {\bibfnamefont {H.~S.}\ \bibnamefont {Reall}},\ }\bibfield  {title} {\enquote {\bibinfo {title} {{Brane world black holes}},}\ }\href {\doibase 10.1103/PhysRevD.61.065007} {\bibfield  {journal} {\bibinfo  {journal} {Phys. Rev. D}\ }\textbf {\bibinfo {volume} {61}},\ \bibinfo {pages} {065007} (\bibinfo {year} {2000})},\ \Eprint {http://arxiv.org/abs/hep-th/9909205} {arXiv:hep-th/9909205} \BibitemShut {NoStop}%
\bibitem [{\citenamefont {Kehagias}\ and\ \citenamefont {Tamvakis}(2001)}]{Kehagias:2000au}%
  \BibitemOpen
  \bibfield  {author} {\bibinfo {author} {\bibfnamefont {A.}~\bibnamefont {Kehagias}}\ and\ \bibinfo {author} {\bibfnamefont {K.}~\bibnamefont {Tamvakis}},\ }\bibfield  {title} {\enquote {\bibinfo {title} {{Localized gravitons, gauge bosons and chiral fermions in smooth spaces generated by a bounce}},}\ }\href {\doibase 10.1016/S0370-2693(01)00274-X} {\bibfield  {journal} {\bibinfo  {journal} {Phys. Lett. B}\ }\textbf {\bibinfo {volume} {504}},\ \bibinfo {pages} {38--46} (\bibinfo {year} {2001})},\ \Eprint {http://arxiv.org/abs/hep-th/0010112} {arXiv:hep-th/0010112} \BibitemShut {NoStop}%
\bibitem [{\citenamefont {Hendi}\ \emph {et~al.}(2020)\citenamefont {Hendi}, \citenamefont {Riazi},\ and\ \citenamefont {Sajadi}}]{Hendi:2020qkk}%
  \BibitemOpen
  \bibfield  {author} {\bibinfo {author} {\bibfnamefont {S.~H.}\ \bibnamefont {Hendi}}, \bibinfo {author} {\bibfnamefont {N.}~\bibnamefont {Riazi}}, \ and\ \bibinfo {author} {\bibfnamefont {S.~N.}\ \bibnamefont {Sajadi}},\ }\bibfield  {title} {\enquote {\bibinfo {title} {{$Z_2$-symmetric thick brane with a specific warp function}},}\ }\href {\doibase 10.1103/PhysRevD.102.124034} {\bibfield  {journal} {\bibinfo  {journal} {Phys. Rev. D}\ }\textbf {\bibinfo {volume} {102}},\ \bibinfo {pages} {124034} (\bibinfo {year} {2020})},\ \Eprint {http://arxiv.org/abs/2011.11093} {arXiv:2011.11093 [gr-qc]} \BibitemShut {NoStop}%
\bibitem [{\citenamefont {Peyravi}\ \emph {et~al.}(2023)\citenamefont {Peyravi}, \citenamefont {Nazifkar}, \citenamefont {Lobo},\ and\ \citenamefont {Javidan}}]{Peyravi:2022ubf}%
  \BibitemOpen
  \bibfield  {author} {\bibinfo {author} {\bibfnamefont {Marzieh}\ \bibnamefont {Peyravi}}, \bibinfo {author} {\bibfnamefont {Samira}\ \bibnamefont {Nazifkar}}, \bibinfo {author} {\bibfnamefont {Francisco S.~N.}\ \bibnamefont {Lobo}}, \ and\ \bibinfo {author} {\bibfnamefont {Kurosh}\ \bibnamefont {Javidan}},\ }\bibfield  {title} {\enquote {\bibinfo {title} {{Thick branes via higher order field theory models with exponential and power-law tails}},}\ }\href {\doibase 10.1140/epjc/s10052-023-11992-z} {\bibfield  {journal} {\bibinfo  {journal} {Eur. Phys. J. C}\ }\textbf {\bibinfo {volume} {83}},\ \bibinfo {pages} {832} (\bibinfo {year} {2023})},\ \Eprint {http://arxiv.org/abs/2210.17387} {arXiv:2210.17387 [gr-qc]} \BibitemShut {NoStop}%
\bibitem [{\citenamefont {Rosa}\ \emph {et~al.}(2022)\citenamefont {Rosa}, \citenamefont {Lob\~ao},\ and\ \citenamefont {Bazeia}}]{Rosa:2022fhl}%
  \BibitemOpen
  \bibfield  {author} {\bibinfo {author} {\bibfnamefont {Jo\~ao~Lu\'\i{}s}\ \bibnamefont {Rosa}}, \bibinfo {author} {\bibfnamefont {A.~S.}\ \bibnamefont {Lob\~ao}}, \ and\ \bibinfo {author} {\bibfnamefont {D.}~\bibnamefont {Bazeia}},\ }\bibfield  {title} {\enquote {\bibinfo {title} {{Impact of compactlike and asymmetric configurations of thick branes on the scalar\textendash{}tensor representation of $f\left( R,T\right) $ gravity}},}\ }\href {\doibase 10.1140/epjc/s10052-022-10159-6} {\bibfield  {journal} {\bibinfo  {journal} {Eur. Phys. J. C}\ }\textbf {\bibinfo {volume} {82}},\ \bibinfo {pages} {191} (\bibinfo {year} {2022})},\ \Eprint {http://arxiv.org/abs/2202.10713} {arXiv:2202.10713 [gr-qc]} \BibitemShut {NoStop}%
\bibitem [{\citenamefont {Shiromizu}\ \emph {et~al.}(2000)\citenamefont {Shiromizu}, \citenamefont {Maeda},\ and\ \citenamefont {Sasaki}}]{Shiromizu:1999wj}%
  \BibitemOpen
  \bibfield  {author} {\bibinfo {author} {\bibfnamefont {Tetsuya}\ \bibnamefont {Shiromizu}}, \bibinfo {author} {\bibfnamefont {Kei-ichi}\ \bibnamefont {Maeda}}, \ and\ \bibinfo {author} {\bibfnamefont {Misao}\ \bibnamefont {Sasaki}},\ }\bibfield  {title} {\enquote {\bibinfo {title} {{The Einstein equation on the 3-brane world}},}\ }\href {\doibase 10.1103/PhysRevD.62.024012} {\bibfield  {journal} {\bibinfo  {journal} {Phys. Rev. D}\ }\textbf {\bibinfo {volume} {62}},\ \bibinfo {pages} {024012} (\bibinfo {year} {2000})},\ \Eprint {http://arxiv.org/abs/gr-qc/9910076} {arXiv:gr-qc/9910076} \BibitemShut {NoStop}%
\bibitem [{\citenamefont {Neves}(2021)}]{Neves:2021dqx}%
  \BibitemOpen
  \bibfield  {author} {\bibinfo {author} {\bibfnamefont {Juliano C.~S.}\ \bibnamefont {Neves}},\ }\bibfield  {title} {\enquote {\bibinfo {title} {{Five-dimensional regular black holes in a brane world}},}\ }\href {\doibase 10.1103/PhysRevD.104.084019} {\bibfield  {journal} {\bibinfo  {journal} {Phys. Rev. D}\ }\textbf {\bibinfo {volume} {104}},\ \bibinfo {pages} {084019} (\bibinfo {year} {2021})},\ \Eprint {http://arxiv.org/abs/2107.04072} {arXiv:2107.04072 [hep-th]} \BibitemShut {NoStop}%
\bibitem [{\citenamefont {Kanti}\ \emph {et~al.}(2002)\citenamefont {Kanti}, \citenamefont {Olasagasti},\ and\ \citenamefont {Tamvakis}}]{Kanti:2002fx}%
  \BibitemOpen
  \bibfield  {author} {\bibinfo {author} {\bibfnamefont {P.}~\bibnamefont {Kanti}}, \bibinfo {author} {\bibfnamefont {I.}~\bibnamefont {Olasagasti}}, \ and\ \bibinfo {author} {\bibfnamefont {K.}~\bibnamefont {Tamvakis}},\ }\bibfield  {title} {\enquote {\bibinfo {title} {{Schwarzschild black branes and strings in higher dimensional brane worlds}},}\ }\href {\doibase 10.1103/PhysRevD.66.104026} {\bibfield  {journal} {\bibinfo  {journal} {Phys. Rev. D}\ }\textbf {\bibinfo {volume} {66}},\ \bibinfo {pages} {104026} (\bibinfo {year} {2002})},\ \Eprint {http://arxiv.org/abs/hep-th/0207283} {arXiv:hep-th/0207283} \BibitemShut {NoStop}%
\bibitem [{\citenamefont {Hirayama}\ and\ \citenamefont {Kang}(2001)}]{Hirayama:2001bi}%
  \BibitemOpen
  \bibfield  {author} {\bibinfo {author} {\bibfnamefont {Takayuki}\ \bibnamefont {Hirayama}}\ and\ \bibinfo {author} {\bibfnamefont {Gungwon}\ \bibnamefont {Kang}},\ }\bibfield  {title} {\enquote {\bibinfo {title} {{Stable black strings in anti-de Sitter space}},}\ }\href {\doibase 10.1103/PhysRevD.64.064010} {\bibfield  {journal} {\bibinfo  {journal} {Phys. Rev. D}\ }\textbf {\bibinfo {volume} {64}},\ \bibinfo {pages} {064010} (\bibinfo {year} {2001})},\ \Eprint {http://arxiv.org/abs/hep-th/0104213} {arXiv:hep-th/0104213} \BibitemShut {NoStop}%
\bibitem [{\citenamefont {Kanti}\ \emph {et~al.}(2003)\citenamefont {Kanti}, \citenamefont {Olasagasti},\ and\ \citenamefont {Tamvakis}}]{Kanti:2003uv}%
  \BibitemOpen
  \bibfield  {author} {\bibinfo {author} {\bibfnamefont {P.}~\bibnamefont {Kanti}}, \bibinfo {author} {\bibfnamefont {I.}~\bibnamefont {Olasagasti}}, \ and\ \bibinfo {author} {\bibfnamefont {K.}~\bibnamefont {Tamvakis}},\ }\bibfield  {title} {\enquote {\bibinfo {title} {{Quest for localized 4-D black holes in brane worlds. 2. Removing the bulk singularities}},}\ }\href {\doibase 10.1103/PhysRevD.68.124001} {\bibfield  {journal} {\bibinfo  {journal} {Phys. Rev. D}\ }\textbf {\bibinfo {volume} {68}},\ \bibinfo {pages} {124001} (\bibinfo {year} {2003})},\ \Eprint {http://arxiv.org/abs/hep-th/0307201} {arXiv:hep-th/0307201} \BibitemShut {NoStop}%
\bibitem [{\citenamefont {Nakas}\ and\ \citenamefont {Kanti}(2021{\natexlab{a}})}]{Nakas:2020sey}%
  \BibitemOpen
  \bibfield  {author} {\bibinfo {author} {\bibfnamefont {Theodoros}\ \bibnamefont {Nakas}}\ and\ \bibinfo {author} {\bibfnamefont {Panagiota}\ \bibnamefont {Kanti}},\ }\bibfield  {title} {\enquote {\bibinfo {title} {{Localized brane-world black hole analytically connected to an AdS$_5$ boundary}},}\ }\href {\doibase 10.1016/j.physletb.2021.136278} {\bibfield  {journal} {\bibinfo  {journal} {Phys. Lett. B}\ }\textbf {\bibinfo {volume} {816}},\ \bibinfo {pages} {136278} (\bibinfo {year} {2021}{\natexlab{a}})},\ \Eprint {http://arxiv.org/abs/2012.09199} {arXiv:2012.09199 [hep-th]} \BibitemShut {NoStop}%
\bibitem [{\citenamefont {Nakas}\ and\ \citenamefont {Kanti}(2021{\natexlab{b}})}]{Nakas:2021srr}%
  \BibitemOpen
  \bibfield  {author} {\bibinfo {author} {\bibfnamefont {Theodoros}\ \bibnamefont {Nakas}}\ and\ \bibinfo {author} {\bibfnamefont {Panagiota}\ \bibnamefont {Kanti}},\ }\bibfield  {title} {\enquote {\bibinfo {title} {{Analytic and exponentially localized braneworld Reissner-Nordstr\"om-AdS solution: A top-down approach}},}\ }\href {\doibase 10.1103/PhysRevD.104.104037} {\bibfield  {journal} {\bibinfo  {journal} {Phys. Rev. D}\ }\textbf {\bibinfo {volume} {104}},\ \bibinfo {pages} {104037} (\bibinfo {year} {2021}{\natexlab{b}})},\ \Eprint {http://arxiv.org/abs/2105.06915} {arXiv:2105.06915 [hep-th]} \BibitemShut {NoStop}%
\bibitem [{\citenamefont {Nakas}\ \emph {et~al.}(2024)\citenamefont {Nakas}, \citenamefont {Pappas},\ and\ \citenamefont {Stuchl\'\i{}k}}]{Nakas:2023yhj}%
  \BibitemOpen
  \bibfield  {author} {\bibinfo {author} {\bibfnamefont {Theodoros}\ \bibnamefont {Nakas}}, \bibinfo {author} {\bibfnamefont {Thomas~D.}\ \bibnamefont {Pappas}}, \ and\ \bibinfo {author} {\bibfnamefont {Zden\v{e}k}\ \bibnamefont {Stuchl\'\i{}k}},\ }\bibfield  {title} {\enquote {\bibinfo {title} {{Bridging dimensions: General embedding algorithm and field-theory reconstruction in 5D braneworld models}},}\ }\href {\doibase 10.1103/PhysRevD.109.L041501} {\bibfield  {journal} {\bibinfo  {journal} {Phys. Rev. D}\ }\textbf {\bibinfo {volume} {109}},\ \bibinfo {pages} {L041501} (\bibinfo {year} {2024})},\ \Eprint {http://arxiv.org/abs/2309.00873} {arXiv:2309.00873 [gr-qc]} \BibitemShut {NoStop}%
\bibitem [{\citenamefont {Crispim}\ \emph {et~al.}(2024{\natexlab{a}})\citenamefont {Crispim}, \citenamefont {Alencar},\ and\ \citenamefont {Estrada}}]{Crispim:2024nou}%
  \BibitemOpen
  \bibfield  {author} {\bibinfo {author} {\bibfnamefont {T.~M.}\ \bibnamefont {Crispim}}, \bibinfo {author} {\bibfnamefont {G.}~\bibnamefont {Alencar}}, \ and\ \bibinfo {author} {\bibfnamefont {Milko}\ \bibnamefont {Estrada}},\ }\bibfield  {title} {\enquote {\bibinfo {title} {{Braneworld Black Bounce to Transversable Wormhole Analytically Connected to an asymptotically $AdS_5$ Boundary}},}\ }\href@noop {} {\  (\bibinfo {year} {2024}{\natexlab{a}})},\ \Eprint {http://arxiv.org/abs/2407.03528} {arXiv:2407.03528 [gr-qc]} \BibitemShut {NoStop}%
\bibitem [{\citenamefont {Crispim}\ \emph {et~al.}(2024{\natexlab{b}})\citenamefont {Crispim}, \citenamefont {Estrada}, \citenamefont {Muniz},\ and\ \citenamefont {Alencar}}]{Crispim:2024yjz}%
  \BibitemOpen
  \bibfield  {author} {\bibinfo {author} {\bibfnamefont {Tiago~M.}\ \bibnamefont {Crispim}}, \bibinfo {author} {\bibfnamefont {Milko}\ \bibnamefont {Estrada}}, \bibinfo {author} {\bibfnamefont {C.~R.}\ \bibnamefont {Muniz}}, \ and\ \bibinfo {author} {\bibfnamefont {G.}~\bibnamefont {Alencar}},\ }\bibfield  {title} {\enquote {\bibinfo {title} {{Braneworld Black Bounce to Transversable Wormhole}},}\ }\href@noop {} {\  (\bibinfo {year} {2024}{\natexlab{b}})},\ \Eprint {http://arxiv.org/abs/2405.08048} {arXiv:2405.08048 [hep-th]} \BibitemShut {NoStop}%
\bibitem [{\citenamefont {Ovalle}(2017)}]{Ovalle:2017fgl}%
  \BibitemOpen
  \bibfield  {author} {\bibinfo {author} {\bibfnamefont {Jorge}\ \bibnamefont {Ovalle}},\ }\bibfield  {title} {\enquote {\bibinfo {title} {{Decoupling gravitational sources in general relativity: from perfect to anisotropic fluids}},}\ }\href {\doibase 10.1103/PhysRevD.95.104019} {\bibfield  {journal} {\bibinfo  {journal} {Phys. Rev. D}\ }\textbf {\bibinfo {volume} {95}},\ \bibinfo {pages} {104019} (\bibinfo {year} {2017})},\ \Eprint {http://arxiv.org/abs/1704.05899} {arXiv:1704.05899 [gr-qc]} \BibitemShut {NoStop}%
\bibitem [{\citenamefont {Contreras}\ \emph {et~al.}(2021)\citenamefont {Contreras}, \citenamefont {Ovalle},\ and\ \citenamefont {Casadio}}]{Contreras:2021yxe}%
  \BibitemOpen
  \bibfield  {author} {\bibinfo {author} {\bibfnamefont {E.}~\bibnamefont {Contreras}}, \bibinfo {author} {\bibfnamefont {J.}~\bibnamefont {Ovalle}}, \ and\ \bibinfo {author} {\bibfnamefont {R.}~\bibnamefont {Casadio}},\ }\bibfield  {title} {\enquote {\bibinfo {title} {{Gravitational decoupling for axially symmetric systems and rotating black holes}},}\ }\href {\doibase 10.1103/PhysRevD.103.044020} {\bibfield  {journal} {\bibinfo  {journal} {Phys. Rev. D}\ }\textbf {\bibinfo {volume} {103}},\ \bibinfo {pages} {044020} (\bibinfo {year} {2021})},\ \Eprint {http://arxiv.org/abs/2101.08569} {arXiv:2101.08569 [gr-qc]} \BibitemShut {NoStop}%
\bibitem [{\citenamefont {Ovalle}\ \emph {et~al.}(2021)\citenamefont {Ovalle}, \citenamefont {Casadio}, \citenamefont {Contreras},\ and\ \citenamefont {Sotomayor}}]{Ovalle:2020kpd}%
  \BibitemOpen
  \bibfield  {author} {\bibinfo {author} {\bibfnamefont {J.}~\bibnamefont {Ovalle}}, \bibinfo {author} {\bibfnamefont {R.}~\bibnamefont {Casadio}}, \bibinfo {author} {\bibfnamefont {E.}~\bibnamefont {Contreras}}, \ and\ \bibinfo {author} {\bibfnamefont {A.}~\bibnamefont {Sotomayor}},\ }\bibfield  {title} {\enquote {\bibinfo {title} {{Hairy black holes by gravitational decoupling}},}\ }\href {\doibase 10.1016/j.dark.2020.100744} {\bibfield  {journal} {\bibinfo  {journal} {Phys. Dark Univ.}\ }\textbf {\bibinfo {volume} {31}},\ \bibinfo {pages} {100744} (\bibinfo {year} {2021})},\ \Eprint {http://arxiv.org/abs/2006.06735} {arXiv:2006.06735 [gr-qc]} \BibitemShut {NoStop}%
\bibitem [{\citenamefont {Ovalle}\ \emph {et~al.}(2018{\natexlab{a}})\citenamefont {Ovalle}, \citenamefont {Casadio}, \citenamefont {Rocha}, \citenamefont {Sotomayor},\ and\ \citenamefont {Stuchlik}}]{Ovalle:2018umz}%
  \BibitemOpen
  \bibfield  {author} {\bibinfo {author} {\bibfnamefont {J.}~\bibnamefont {Ovalle}}, \bibinfo {author} {\bibfnamefont {R.}~\bibnamefont {Casadio}}, \bibinfo {author} {\bibfnamefont {R.~da}\ \bibnamefont {Rocha}}, \bibinfo {author} {\bibfnamefont {A.}~\bibnamefont {Sotomayor}}, \ and\ \bibinfo {author} {\bibfnamefont {Z.}~\bibnamefont {Stuchlik}},\ }\bibfield  {title} {\enquote {\bibinfo {title} {{Black holes by gravitational decoupling}},}\ }\href {\doibase 10.1140/epjc/s10052-018-6450-4} {\bibfield  {journal} {\bibinfo  {journal} {Eur. Phys. J. C}\ }\textbf {\bibinfo {volume} {78}},\ \bibinfo {pages} {960} (\bibinfo {year} {2018}{\natexlab{a}})},\ \Eprint {http://arxiv.org/abs/1804.03468} {arXiv:1804.03468 [gr-qc]} \BibitemShut {NoStop}%
\bibitem [{\citenamefont {Ovalle}\ \emph {et~al.}(2023)\citenamefont {Ovalle}, \citenamefont {Casadio},\ and\ \citenamefont {Giusti}}]{Ovalle:2023ref}%
  \BibitemOpen
  \bibfield  {author} {\bibinfo {author} {\bibfnamefont {Jorge}\ \bibnamefont {Ovalle}}, \bibinfo {author} {\bibfnamefont {Roberto}\ \bibnamefont {Casadio}}, \ and\ \bibinfo {author} {\bibfnamefont {Andrea}\ \bibnamefont {Giusti}},\ }\bibfield  {title} {\enquote {\bibinfo {title} {{Regular hairy black holes through Minkowski deformation}},}\ }\href {\doibase 10.1016/j.physletb.2023.138085} {\bibfield  {journal} {\bibinfo  {journal} {Phys. Lett. B}\ }\textbf {\bibinfo {volume} {844}},\ \bibinfo {pages} {138085} (\bibinfo {year} {2023})},\ \Eprint {http://arxiv.org/abs/2304.03263} {arXiv:2304.03263 [gr-qc]} \BibitemShut {NoStop}%
\bibitem [{\citenamefont {Estrada}\ and\ \citenamefont {Tello-Ortiz}(2018)}]{Estrada:2018zbh}%
  \BibitemOpen
  \bibfield  {author} {\bibinfo {author} {\bibfnamefont {Milko}\ \bibnamefont {Estrada}}\ and\ \bibinfo {author} {\bibfnamefont {Francisco}\ \bibnamefont {Tello-Ortiz}},\ }\bibfield  {title} {\enquote {\bibinfo {title} {{A new family of analytical anisotropic solutions by gravitational decoupling}},}\ }\href {\doibase 10.1140/epjp/i2018-12249-9} {\bibfield  {journal} {\bibinfo  {journal} {Eur. Phys. J. Plus}\ }\textbf {\bibinfo {volume} {133}},\ \bibinfo {pages} {453} (\bibinfo {year} {2018})},\ \Eprint {http://arxiv.org/abs/1803.02344} {arXiv:1803.02344 [gr-qc]} \BibitemShut {NoStop}%
\bibitem [{\citenamefont {Estrada}\ and\ \citenamefont {Prado}(2019)}]{Estrada:2018vrl}%
  \BibitemOpen
  \bibfield  {author} {\bibinfo {author} {\bibfnamefont {Milko}\ \bibnamefont {Estrada}}\ and\ \bibinfo {author} {\bibfnamefont {Reginaldo}\ \bibnamefont {Prado}},\ }\bibfield  {title} {\enquote {\bibinfo {title} {{The Gravitational decoupling method: the higher dimensional case to find new analytic solutions}},}\ }\href {\doibase 10.1140/epjp/i2019-12555-8} {\bibfield  {journal} {\bibinfo  {journal} {Eur. Phys. J. Plus}\ }\textbf {\bibinfo {volume} {134}},\ \bibinfo {pages} {168} (\bibinfo {year} {2019})},\ \Eprint {http://arxiv.org/abs/1809.03591} {arXiv:1809.03591 [gr-qc]} \BibitemShut {NoStop}%
\bibitem [{\citenamefont {Sharif}\ and\ \citenamefont {Sadiq}(2018)}]{Sharif:2018toc}%
  \BibitemOpen
  \bibfield  {author} {\bibinfo {author} {\bibfnamefont {M.}~\bibnamefont {Sharif}}\ and\ \bibinfo {author} {\bibfnamefont {Sobia}\ \bibnamefont {Sadiq}},\ }\bibfield  {title} {\enquote {\bibinfo {title} {{Gravitational Decoupled Charged Anisotropic Spherical Solutions}},}\ }\href {\doibase 10.1140/epjc/s10052-018-5894-x} {\bibfield  {journal} {\bibinfo  {journal} {Eur. Phys. J. C}\ }\textbf {\bibinfo {volume} {78}},\ \bibinfo {pages} {410} (\bibinfo {year} {2018})},\ \Eprint {http://arxiv.org/abs/1804.09616} {arXiv:1804.09616 [gr-qc]} \BibitemShut {NoStop}%
\bibitem [{\citenamefont {Estrada}(2019)}]{Estrada:2019aeh}%
  \BibitemOpen
  \bibfield  {author} {\bibinfo {author} {\bibfnamefont {Milko}\ \bibnamefont {Estrada}},\ }\bibfield  {title} {\enquote {\bibinfo {title} {{A way of decoupling gravitational sources in pure Lovelock gravity}},}\ }\href {\doibase 10.1140/epjc/s10052-019-7444-6} {\bibfield  {journal} {\bibinfo  {journal} {Eur. Phys. J. C}\ }\textbf {\bibinfo {volume} {79}},\ \bibinfo {pages} {918} (\bibinfo {year} {2019})},\ \bibinfo {note} {[Erratum: Eur.Phys.J.C 80, 590 (2020)]},\ \Eprint {http://arxiv.org/abs/1905.12129} {arXiv:1905.12129 [gr-qc]} \BibitemShut {NoStop}%
\bibitem [{\citenamefont {Le\'on}\ and\ \citenamefont {Sotomayor}(2019)}]{Leon:2019abq}%
  \BibitemOpen
  \bibfield  {author} {\bibinfo {author} {\bibfnamefont {P.}~\bibnamefont {Le\'on}}\ and\ \bibinfo {author} {\bibfnamefont {A.}~\bibnamefont {Sotomayor}},\ }\bibfield  {title} {\enquote {\bibinfo {title} {{Braneworld Gravity under gravitational decoupling}},}\ }\href {\doibase 10.1002/prop.201900077} {\bibfield  {journal} {\bibinfo  {journal} {Fortsch. Phys.}\ }\textbf {\bibinfo {volume} {67}},\ \bibinfo {pages} {1900077} (\bibinfo {year} {2019})},\ \Eprint {http://arxiv.org/abs/1907.11763} {arXiv:1907.11763 [gr-qc]} \BibitemShut {NoStop}%
\bibitem [{\citenamefont {Ovalle}\ \emph {et~al.}(2018{\natexlab{b}})\citenamefont {Ovalle}, \citenamefont {Casadio}, \citenamefont {da~Rocha},\ and\ \citenamefont {Sotomayor}}]{Ovalle:2017wqi}%
  \BibitemOpen
  \bibfield  {author} {\bibinfo {author} {\bibfnamefont {J.}~\bibnamefont {Ovalle}}, \bibinfo {author} {\bibfnamefont {R.}~\bibnamefont {Casadio}}, \bibinfo {author} {\bibfnamefont {R.}~\bibnamefont {da~Rocha}}, \ and\ \bibinfo {author} {\bibfnamefont {A.}~\bibnamefont {Sotomayor}},\ }\bibfield  {title} {\enquote {\bibinfo {title} {{Anisotropic solutions by gravitational decoupling}},}\ }\href {\doibase 10.1140/epjc/s10052-018-5606-6} {\bibfield  {journal} {\bibinfo  {journal} {Eur. Phys. J. C}\ }\textbf {\bibinfo {volume} {78}},\ \bibinfo {pages} {122} (\bibinfo {year} {2018}{\natexlab{b}})},\ \Eprint {http://arxiv.org/abs/1708.00407} {arXiv:1708.00407 [gr-qc]} \BibitemShut {NoStop}%
\bibitem [{\citenamefont {Sharif}\ and\ \citenamefont {Saba}(2020)}]{Sharif:2020llo}%
  \BibitemOpen
  \bibfield  {author} {\bibinfo {author} {\bibfnamefont {M.}~\bibnamefont {Sharif}}\ and\ \bibinfo {author} {\bibfnamefont {Saadia}\ \bibnamefont {Saba}},\ }\bibfield  {title} {\enquote {\bibinfo {title} {{Gravitational decoupled Durgapal\textendash{}Fuloria anisotropic solutions in modified Gauss\textendash{}Bonnet gravity}},}\ }\href {\doibase 10.1016/j.cjph.2019.11.023} {\bibfield  {journal} {\bibinfo  {journal} {Chin. J. Phys.}\ }\textbf {\bibinfo {volume} {63}},\ \bibinfo {pages} {348--364} (\bibinfo {year} {2020})}\BibitemShut {NoStop}%
\bibitem [{\citenamefont {da~Rocha}(2022)}]{daRocha:2021sqd}%
  \BibitemOpen
  \bibfield  {author} {\bibinfo {author} {\bibfnamefont {Roldao}\ \bibnamefont {da~Rocha}},\ }\bibfield  {title} {\enquote {\bibinfo {title} {{Gravitational decoupling of generalized Horndeski hybrid stars}},}\ }\href {\doibase 10.1140/epjc/s10052-021-09971-3} {\bibfield  {journal} {\bibinfo  {journal} {Eur. Phys. J. C}\ }\textbf {\bibinfo {volume} {82}},\ \bibinfo {pages} {34} (\bibinfo {year} {2022})},\ \Eprint {http://arxiv.org/abs/2111.11995} {arXiv:2111.11995 [gr-qc]} \BibitemShut {NoStop}%
\bibitem [{\citenamefont {Kanti}\ and\ \citenamefont {Tamvakis}(2002)}]{Kanti:2001cj}%
  \BibitemOpen
  \bibfield  {author} {\bibinfo {author} {\bibfnamefont {Panagiota}\ \bibnamefont {Kanti}}\ and\ \bibinfo {author} {\bibfnamefont {Kyriakos}\ \bibnamefont {Tamvakis}},\ }\bibfield  {title} {\enquote {\bibinfo {title} {{Quest for localized 4-D black holes in brane worlds}},}\ }\href {\doibase 10.1103/PhysRevD.65.084010} {\bibfield  {journal} {\bibinfo  {journal} {Phys. Rev. D}\ }\textbf {\bibinfo {volume} {65}},\ \bibinfo {pages} {084010} (\bibinfo {year} {2002})},\ \Eprint {http://arxiv.org/abs/hep-th/0110298} {arXiv:hep-th/0110298} \BibitemShut {NoStop}%
\bibitem [{\citenamefont {Binetruy}\ \emph {et~al.}(2000)\citenamefont {Binetruy}, \citenamefont {Deffayet},\ and\ \citenamefont {Langlois}}]{Binetruy:1999ut}%
  \BibitemOpen
  \bibfield  {author} {\bibinfo {author} {\bibfnamefont {Pierre}\ \bibnamefont {Binetruy}}, \bibinfo {author} {\bibfnamefont {Cedric}\ \bibnamefont {Deffayet}}, \ and\ \bibinfo {author} {\bibfnamefont {David}\ \bibnamefont {Langlois}},\ }\bibfield  {title} {\enquote {\bibinfo {title} {{Nonconventional cosmology from a brane universe}},}\ }\href {\doibase 10.1016/S0550-3213(99)00696-3} {\bibfield  {journal} {\bibinfo  {journal} {Nucl. Phys. B}\ }\textbf {\bibinfo {volume} {565}},\ \bibinfo {pages} {269--287} (\bibinfo {year} {2000})},\ \Eprint {http://arxiv.org/abs/hep-th/9905012} {arXiv:hep-th/9905012} \BibitemShut {NoStop}%
\bibitem [{\citenamefont {Alencar}\ \emph {et~al.}(2024)\citenamefont {Alencar}, \citenamefont {Gogberashvili},\ and\ \citenamefont {Costa~Filho}}]{Alencar:2024lrl}%
  \BibitemOpen
  \bibfield  {author} {\bibinfo {author} {\bibfnamefont {G.}~\bibnamefont {Alencar}}, \bibinfo {author} {\bibfnamefont {M.}~\bibnamefont {Gogberashvili}}, \ and\ \bibinfo {author} {\bibfnamefont {R.~N.}\ \bibnamefont {Costa~Filho}},\ }\bibfield  {title} {\enquote {\bibinfo {title} {{Local Sum Rules for 5D Braneworlds}},}\ }\href@noop {} {\  (\bibinfo {year} {2024})},\ \Eprint {http://arxiv.org/abs/2405.04567} {arXiv:2405.04567 [hep-ph]} \BibitemShut {NoStop}%
\bibitem [{\citenamefont {Hayward}(2006)}]{Hayward:2005gi}%
  \BibitemOpen
  \bibfield  {author} {\bibinfo {author} {\bibfnamefont {Sean~A.}\ \bibnamefont {Hayward}},\ }\bibfield  {title} {\enquote {\bibinfo {title} {{Formation and evaporation of regular black holes}},}\ }\href {\doibase 10.1103/PhysRevLett.96.031103} {\bibfield  {journal} {\bibinfo  {journal} {Phys. Rev. Lett.}\ }\textbf {\bibinfo {volume} {96}},\ \bibinfo {pages} {031103} (\bibinfo {year} {2006})},\ \Eprint {http://arxiv.org/abs/gr-qc/0506126} {arXiv:gr-qc/0506126} \BibitemShut {NoStop}%
\bibitem [{\citenamefont {Dymnikova}(1992)}]{Dymnikova:1992ux}%
  \BibitemOpen
  \bibfield  {author} {\bibinfo {author} {\bibfnamefont {I.}~\bibnamefont {Dymnikova}},\ }\bibfield  {title} {\enquote {\bibinfo {title} {{Vacuum nonsingular black hole}},}\ }\href {\doibase 10.1007/BF00760226} {\bibfield  {journal} {\bibinfo  {journal} {Gen. Rel. Grav.}\ }\textbf {\bibinfo {volume} {24}},\ \bibinfo {pages} {235--242} (\bibinfo {year} {1992})}\BibitemShut {NoStop}%
\bibitem [{\citenamefont {Spallucci}\ and\ \citenamefont {Smailagic}(2017)}]{Spallucci:2017aod}%
  \BibitemOpen
  \bibfield  {author} {\bibinfo {author} {\bibfnamefont {Euro}\ \bibnamefont {Spallucci}}\ and\ \bibinfo {author} {\bibfnamefont {Anais}\ \bibnamefont {Smailagic}},\ }\bibfield  {title} {\enquote {\bibinfo {title} {{Regular black holes from semi-classical down to Planckian size}},}\ }\href {\doibase 10.1142/S0218271817300130} {\bibfield  {journal} {\bibinfo  {journal} {Int. J. Mod. Phys. D}\ }\textbf {\bibinfo {volume} {26}},\ \bibinfo {pages} {1730013} (\bibinfo {year} {2017})},\ \Eprint {http://arxiv.org/abs/1701.04592} {arXiv:1701.04592 [hep-th]} \BibitemShut {NoStop}%
\bibitem [{\citenamefont {Gregory}\ and\ \citenamefont {Laflamme}(1993)}]{Gregory:1993vy}%
  \BibitemOpen
  \bibfield  {author} {\bibinfo {author} {\bibfnamefont {R.}~\bibnamefont {Gregory}}\ and\ \bibinfo {author} {\bibfnamefont {R.}~\bibnamefont {Laflamme}},\ }\bibfield  {title} {\enquote {\bibinfo {title} {{Black strings and p-branes are unstable}},}\ }\href {\doibase 10.1103/PhysRevLett.70.2837} {\bibfield  {journal} {\bibinfo  {journal} {Phys. Rev. Lett.}\ }\textbf {\bibinfo {volume} {70}},\ \bibinfo {pages} {2837--2840} (\bibinfo {year} {1993})},\ \Eprint {http://arxiv.org/abs/hep-th/9301052} {arXiv:hep-th/9301052} \BibitemShut {NoStop}%
\bibitem [{\citenamefont {Gregory}(2000)}]{Gregory:2000gf}%
  \BibitemOpen
  \bibfield  {author} {\bibinfo {author} {\bibfnamefont {Ruth}\ \bibnamefont {Gregory}},\ }\bibfield  {title} {\enquote {\bibinfo {title} {{Black string instabilities in Anti-de Sitter space}},}\ }\href {\doibase 10.1088/0264-9381/17/18/103} {\bibfield  {journal} {\bibinfo  {journal} {Class. Quant. Grav.}\ }\textbf {\bibinfo {volume} {17}},\ \bibinfo {pages} {L125--L132} (\bibinfo {year} {2000})},\ \Eprint {http://arxiv.org/abs/hep-th/0004101} {arXiv:hep-th/0004101} \BibitemShut {NoStop}%
\bibitem [{\citenamefont {Germ\'an}\ \emph {et~al.}(2016)\citenamefont {Germ\'an}, \citenamefont {Herrera-Aguilar}, \citenamefont {Kuerten}, \citenamefont {Malagon-Morejon},\ and\ \citenamefont {da~Rocha}}]{German:2015cna}%
  \BibitemOpen
  \bibfield  {author} {\bibinfo {author} {\bibfnamefont {Gabriel}\ \bibnamefont {Germ\'an}}, \bibinfo {author} {\bibfnamefont {Alfredo}\ \bibnamefont {Herrera-Aguilar}}, \bibinfo {author} {\bibfnamefont {Andre~M.}\ \bibnamefont {Kuerten}}, \bibinfo {author} {\bibfnamefont {Dagoberto}\ \bibnamefont {Malagon-Morejon}}, \ and\ \bibinfo {author} {\bibfnamefont {Roldao}\ \bibnamefont {da~Rocha}},\ }\bibfield  {title} {\enquote {\bibinfo {title} {{Stability of a tachyon braneworld}},}\ }\href {\doibase 10.1088/1475-7516/2016/01/047} {\bibfield  {journal} {\bibinfo  {journal} {JCAP}\ }\textbf {\bibinfo {volume} {01}},\ \bibinfo {pages} {047} (\bibinfo {year} {2016})},\ \Eprint {http://arxiv.org/abs/1508.03867} {arXiv:1508.03867 [hep-th]} \BibitemShut {NoStop}%
\bibitem [{\citenamefont {Koley}\ and\ \citenamefont {Kar}(2005)}]{Koley:2005nv}%
  \BibitemOpen
  \bibfield  {author} {\bibinfo {author} {\bibfnamefont {Ratna}\ \bibnamefont {Koley}}\ and\ \bibinfo {author} {\bibfnamefont {Sayan}\ \bibnamefont {Kar}},\ }\bibfield  {title} {\enquote {\bibinfo {title} {{A Novel braneworld model with a bulk scalar field}},}\ }\href {\doibase 10.1016/j.physletb.2005.09.063} {\bibfield  {journal} {\bibinfo  {journal} {Phys. Lett. B}\ }\textbf {\bibinfo {volume} {623}},\ \bibinfo {pages} {244--250} (\bibinfo {year} {2005})},\ \bibinfo {note} {[Erratum: Phys.Lett.B 631, 199 (2005)]},\ \Eprint {http://arxiv.org/abs/hep-th/0507277} {arXiv:hep-th/0507277} \BibitemShut {NoStop}%
\bibitem [{\citenamefont {Casadio}\ and\ \citenamefont {Mazzacurati}(2003)}]{Casadio:2002uv}%
  \BibitemOpen
  \bibfield  {author} {\bibinfo {author} {\bibfnamefont {Roberto}\ \bibnamefont {Casadio}}\ and\ \bibinfo {author} {\bibfnamefont {Lorenzo}\ \bibnamefont {Mazzacurati}},\ }\bibfield  {title} {\enquote {\bibinfo {title} {{Bulk shape of brane world black holes}},}\ }\href {\doibase 10.1142/S0217732303009794} {\bibfield  {journal} {\bibinfo  {journal} {Mod. Phys. Lett. A}\ }\textbf {\bibinfo {volume} {18}},\ \bibinfo {pages} {651--660} (\bibinfo {year} {2003})},\ \Eprint {http://arxiv.org/abs/gr-qc/0205129} {arXiv:gr-qc/0205129} \BibitemShut {NoStop}%
\bibitem [{\citenamefont {Chamblin}\ \emph {et~al.}(2001)\citenamefont {Chamblin}, \citenamefont {Reall}, \citenamefont {Shinkai},\ and\ \citenamefont {Shiromizu}}]{Chamblin:2000ra}%
  \BibitemOpen
  \bibfield  {author} {\bibinfo {author} {\bibfnamefont {Andrew}\ \bibnamefont {Chamblin}}, \bibinfo {author} {\bibfnamefont {Harvey~S.}\ \bibnamefont {Reall}}, \bibinfo {author} {\bibfnamefont {Hisa-aki}\ \bibnamefont {Shinkai}}, \ and\ \bibinfo {author} {\bibfnamefont {Tetsuya}\ \bibnamefont {Shiromizu}},\ }\bibfield  {title} {\enquote {\bibinfo {title} {{Charged brane world black holes}},}\ }\href {\doibase 10.1103/PhysRevD.63.064015} {\bibfield  {journal} {\bibinfo  {journal} {Phys. Rev. D}\ }\textbf {\bibinfo {volume} {63}},\ \bibinfo {pages} {064015} (\bibinfo {year} {2001})},\ \Eprint {http://arxiv.org/abs/hep-th/0008177} {arXiv:hep-th/0008177} \BibitemShut {NoStop}%
\end{thebibliography}%

\end{document}